\UseRawInputEncoding
\documentclass[usenatbib,useAMS,usedcolumn]{mnras}

\usepackage{epsfig}
\usepackage{times,pdflscape}
\pdfoutput=1
\hypersetup{pdfauthor={H.L. Liu},
            bookmarksnumbered=true}
\newcommand{\cm}{\ensuremath{\mbox{~cm}}}
\newcommand{\pcmsq}{\ensuremath{\cm^{-2}}}
\newcommand{\pcmcu}{\ensuremath{\cm^{-3}}}

\newcommand{\vel}{km\,s$^{-1}$}

\newcommand{\msun}{$M_{\odot}$}
\newcommand{\lsun}{$L_{\odot}$}
\newcommand{\um}{$\mu$m}

\newcommand{\egcite}{\citep[e.g.,][]}

\newcommand{\hcop}{HCO$^{+}$}
\newcommand{\htcop}{H$^{13}$CO$^{+}$}
\newcommand{\halpha}{H$40_{\alpha}$}

\newcommand{\chthoh}{CH$_3$OH}

\newcommand{\uchii}{UC-H\textsc{ii}}

\newcommand{\hii}{H\textsc{ii}}

\newcommand{\filname}{G34}

\graphicspath{{./0figs/}}

\begin{document}

\title[Fragmentation and gas dynamics in IRDC G034.43+00.24
] 
{ATOMS: ALMA Three-millimeter Observations of Massive Star-forming regions - VI. Hierarchical fragmentation and gas dynamics in IRDC G034.43+00.24
}

% % If you need two or more lines of authors, add an extra line using \newauthor

\author[H.-L. Liu et al.]{
%Hong-Li Liu,\thanks{E-mail: hongliliu2012@gmail.com (HLL)}$^{1,3}$
%Tie Liu,\thanks{E-mail: liutie@shao.ac.cn (TL)}$^{2}$
%Hong-Li Liu\,\orcidlink{0000-0003-3343-9645},$^{\star 1}$
Hong-Li Liu,$^{\star 1}$
Anandmayee Tej,$^{\star 2}$
Tie Liu,$^{\star 3,4}$
Namitha Issac,$^{2}$
Anindya Saha,$^{2}$
Paul F. Goldsmith,$^{5}$
\newauthor
Jun-Zhi Wang,$^{3,6}$
Qizhou Zhang,$^{7}$
Sheng-Li Qin,$^{1}$
Ke Wang,$^{8,9}$
Shanghuo Li,$^{10}$
Archana Soam,$^{11}$
\newauthor
Lokesh Dewangan,$^{12}$
Chang Won Lee,$^{10,13}$
Pak-Shing Li,$^{14}$
Xun-Chuan Liu,$^{9}$
Yong Zhang,$^{15}$
Zhiyuan Ren,$^{16}$
\newauthor
Mika Juvela,$^{17}$
Leonardo Bronfman,$^{18}$
Yue-Fang Wu,$^{9,8}$
Ken'ichi Tatematsu,$^{19}$
Xi Chen, $^{20}$
Di Li,$^{16,21,22}$
\newauthor
Amelia Stutz,$^{23,24}$
Siju Zhang,$^{8}$
L. Viktor Toth,$^{25}$
Qiu-Yi Luo,$^{3}$
Feng-Wei Xu,$^{8,9}$
Jinzeng Li,$^{16}$
Rong Liu,$^{16}$
\newauthor
Jianwen Zhou,$^{16}$
Chao Zhang,$^{26}$
Mengyao Tang,$^{1}$
Chao Zhang,$^{1}$
Tapas Baug,$^{27}$
E. Mannfors,$^{17}$
\newauthor
Eswaraiah Chakali,$^{16,28}$
Somnath  Dutta,$^{29}$
\\
Affiliations are listed at the end of the paper}

%\date{Accepted 7 May 2021; Received 7 May; in original 22 December 2020}

\pagerange{\pageref{firstpage}--\pageref{lastpage}} \pubyear{2021}

\maketitle

\label{firstpage}

\begin{abstract} 
 We present new 3\,mm continuum and molecular lines observations from the ATOMS survey towards the massive protostellar clump, MM1, located in the  filamentary infrared dark cloud (IRDC), G034.43+00.24 (\filname).
The lines  observed are the tracers of either dense gas (e.g., HCO$^+$/H$^{13}$CO$^+$ J=1--0) or 
outflows (e.g., CS J=2--1). 
The most complete picture to date of seven cores in MM1  is revealed by dust continuum emission.
These cores are  found to be gravitationally bound, with virial parameter,  $\alpha_{\rm vir} < 2$.
At least four outflows are identified in MM1 with a total outflowing mass of $\sim 45$\,\msun, and a total energy of $1\times10^{47}$\,ergs, typical of outflows from a B0-type star.  Evidence of hierarchical fragmentation, where turbulence dominates over thermal pressure, is observed at both the cloud and the clump scales.
This could be linked to the scale-dependent, dynamical mass inflow/accretion on clump and core scales.
We therefore suggest that the \filname\ cloud could be undergoing a dynamical mass inflow/accretion process linked to the multi-scale fragmentation, which leads to the sequential formation of fragments of the initial cloud, clumps, and ultimately dense cores, the sites of star formation.

\end{abstract} 


\footnotetext[1]{E-mail: hongliliu2012@gmail.com, liutie@shao.ac.cn, and tej@iist.ac.in}

\section{Introduction} \label{sec:intro}
High-mass stars ($M_{\star}>8$\,\msun) are of great importance in many astrophysical processes ranging from the production and transfer of heavy elements via nucleosynthesis, to the structure and evolution of their host galaxies, and even to future star formation in their natal 
molecular clouds \egcite{Ken05,Urq13}. 
However, high-mass star formation remains poorly understood due to the observational challenges stemming from the relatively large distances to young high-mass stars, their rarity, opaque surroundings, short lifetime, and the complicated, crowded cluster environment \egcite{Zin07,Tan14,Mot18}. 
High-mass  stars are known to form  mainly in clusters through hierarchical fragmentation.
In both observations and theoretical treatments \egcite{Zha09,Wan11,Per13, Wan14,Beu18,Yua18,Mot18,Vaz19}, hierarchical fragmentation has been observed to proceed
 on almost all scales from cloud, through filaments and clumps, to individual star-forming cores\footnote{The nomenclature of \citet{Zha09} and \citet{Wan11, Wan14} is adopted where a cloud is referred to as a structure of $>1$\,pc size, a clump as a structure of $\sim1$\,pc size, and a core as a structure of $\sim$0.1\,pc size. A core does not necessarily collapse into a single star but can fragment into substructures (a.k.a., condensations) and form a small cluster of stars.}, ultimately leading to a cluster of young stars. Moreover, the entire fragmentation process has been recognized to play a crucial role in determining the final mass of the individual stars formed, and thus the initial mass function, where the latter is a key input to the current theories.

Several efforts have been made to investigate the  process of fragmentation, especially on clump and core scales \egcite{Pal13,Pal14,Cse17,Beu18,Svo19,San19,Li19,Li20,Pal21}. 
However, finding different degrees of fragmentation makes it difficult to draw a decisive conclusion about the modality of fragmentation, which is a key parameter, especially for high-mass star formation. 
For example,  \citet{Cse17} found a low level of fragmentation with fragment masses above 40\,\msun\ in their ALMA-880\,\um\ (i.e., 340.1\,GHz) observations of 35 massive clumps down to 0.06\,pc scale.
In contrast, based on ALMA-1\,mm observations towards 12 infrared dark cloud (IRDC) clumps, 
\citet{San19} revealed a higher level of fragmentation with a large population of low-mass ($\leq1$\,\msun) cores  of sizes $\leq0.1$\,pc but no high-mass  counterparts ($\geq30$\,\msun). 
These different degrees of fragmentation probably represent different modalities that control the mass reservoir feeding the individual
stars, as predicted  by the two main competing theories  of high-mass star formation: ``core-accretion'' \citep{McK03} and ``competitive-accretion'' \citep{Bon04}.
The core-accretion hypothesis, which is essentially a scaled-up version of low-mass star formation, favors the fragmentation of a clump into massive cores that proceed to form high-mass stars. In comparison, the competitive-accretion theory predicts the fragmentation of a clump into a larger population of low-mass cores that competitively accrete from the common mass reservoir. In this framework, cores located at the gravitational well of the system preferentially form high-mass stars.
Therefore, more detailed investigation is still needed to establish the link between the  observations of the degree of fragmentation and theoretical predictions.

Fragmentation  tends to be associated with rich and  complex kinematics and dynamics as a result of  the interaction among gravity, turbulence, magnetic fields, and/or other factors such as intense radiative feedback from newly-formed stars.
Both kinematics and dynamics are therefore thought to be   useful probes for dissecting  the underlying physics (e.g. mass accretion, outflows)  related to hierarchical, multi-scale fragmentation processes in star formation.
Recent state-of-the-art numerical simulations of cloud complexes \egcite{Pad20} have reproduced a web of filamentary structures, each with longitudinal velocity gradient indicative of a mass flow along the filament converging towards the web node, where high-mass young stellar objects (YSOs) are preferentially found to  reside.  
The mass of the final stars has  therefore been suggested to be 
regulated not only by the clump- or core-scale mass accretion but also by the larger-scale mass inflow/accretion. 
 In fact, this multi-scale mass inflow/accretion has been revealed in previous multi-scale kinematic observations \citep{Zha11,Per13,Che17,Yua18}.
For example,  in their study of the high-mass protostellar clump, G22, \citet{Yua18} observed that the protostar grows in mass simultaneously via core, clump, and cloud-fed accretion with an increasing trend of mass inflow/infall rates of $7.4\times10^{-5}$\,$M_\odot\rm yr^{-1}$, $7.2\times10^{-4}$\,$M_\odot\rm yr^{-1}$,
 and $\sim 100$\,$M_\odot\rm yr^{-1}$, respectively. 
 It thus appears extremely promising to conduct similar studies to obtain an in-depth understanding of the kinematics and dynamics involved in the process of multi-scale fragmentation associated with high-mass star formation.
 
\begin{figure*}
\centering
\includegraphics[width=2.7 in]{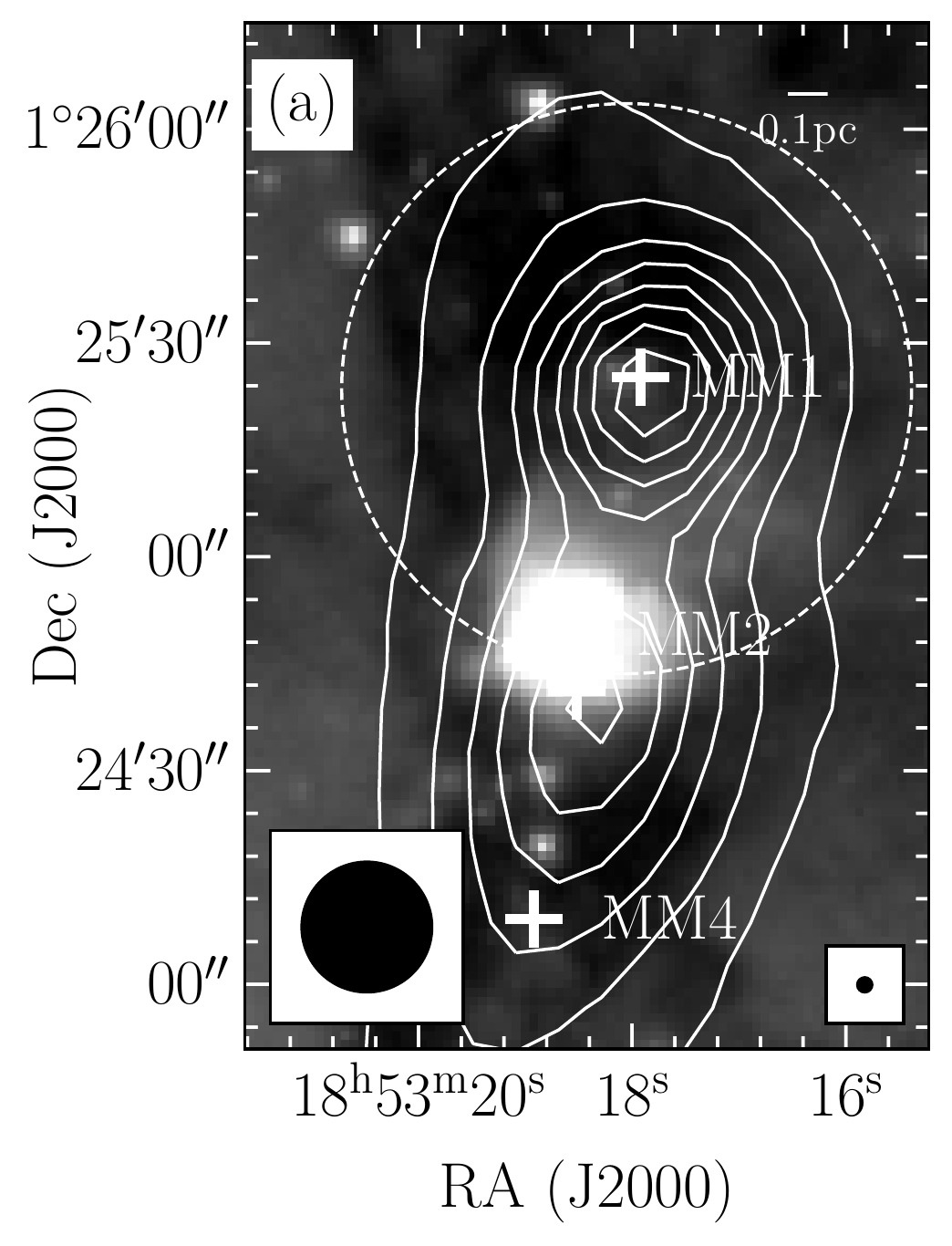}
\includegraphics[width=3.7 in]{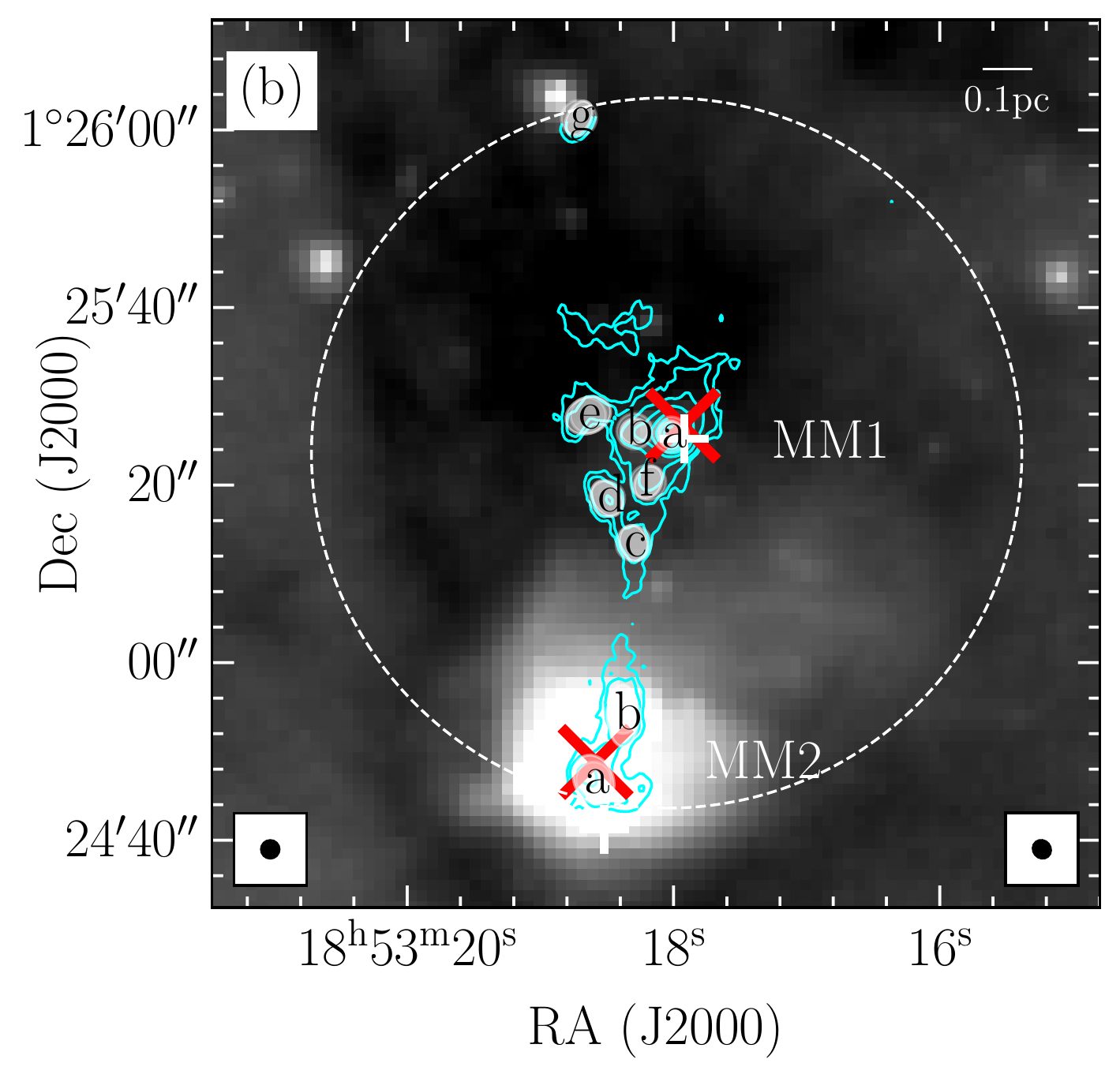}
\caption{ 
(a) The two massive protostellar clumps, MM1 and MM2, of IRDC \filname\ in {\it Spitzer}  8.0\,$\mu$m image overlaid with the ATLASGAL 870\,$\mu$m continuum contours. 
The contours start at 3\,rms (rms~$\sim0.6$~Jy beam$^{-1}$) increasing in steps following the power-law $D=3\times N^{p}+2$, where $D$ is the dynamical range of the intensity map (i.e., the ratio between the peak and the rms noise), $N$ is the number of contours used (8 in this case).
The millimeter clumps identified from the 1.2\,mm continuum  maps presented by \citet{Rat06} are indicated  with plus symbols. 
The beams of the 8.0\,$\mu$m and 870\,$\mu$m data are  shown at the bottom left, and bottom right corners, respectively.
(b)  Close-up view of MM1 and MM2. 
The contours represent the ATOMS 3\,mm continuum  emission. Note that the ATOMS map is not corrected for the primary beam.  The contour levels start at
 3\,rms (rms~$\sim0.3$~mJy~beam$^{-1}$) with steps following the same power-law form as in panel\,(a). Labels a--g identify the cores  extracted from the continuum emission. The beams of the 8.0\,$\mu$m and 3\,mm data are  shown at the bottom left, and bottom right corners, respectively.
 The locations of the B2-type YSO associated with MM1 and the \uchii\ region associated with MM2 are indicated with red X symbols. The dashed circle in both panels  demarcates the field view of the ATOMS $3$\,mm observations.}
\label{fig:overview}
\end{figure*}

The primary target for this study is the massive clump, MM1, located in  the well-known  filamentary IRDC, G034.43+00.24 (hereafter G34 \citealt{Rat05,Rat06,Lu 14,Liu20a}). 
We have also probed another associated massive clump, MM2, which is partly covered with the ATOMS survey.
 We adopt a kinematic distance of $3.7\pm 0.3$\,kpc \egcite{Rat05, Xu 16, Tan19, Liu20a}  to this cloud. 
The two massive clumps have a few hundred solar masses within diameters of 0.2--0.5\,pc.
MM1 has a bolometric luminosity of $2.4\times10^4$\,\lsun\ mainly  due to an associated B0-type YSO (Fig.\,\ref{fig:overview}), while
MM2 has  a bolometric luminosity of $1.4\times10^4$\,\lsun\ mainly from an associated ultra-compact \hii\ region (\uchii, i.e., IRAS\,18507+0121, \citealt{Bro96}). 
The coexistence of these star-forming signatures and the IRDC nature of the \filname\ cloud suggests early stages of high-mass star formation in the clumps, MM1 and MM2.

In this paper, we present our new ALMA 3\,mm observations towards IRDC\,G34, which is part of the ATOMS survey\footnote{ATOMS: ALMA Three-millimeter Observations of Massive Star-forming regions survey} (Project ID: 2019.1.00685.S,  \citealt{Liu20b,Liu20c,Liu21}, hereafter Paper\,I, Paper\,II, and Paper\,III, respectively, see Sect.\,\ref{sec:observations}). 
The  ATOMS survey is aimed to  investigate statistically the relation between high-mass star formation and 
the distribution of dense gas, filamentary structures, and feedback, by observing 3\,mm continuum and gas emission at a nearly uniform angular resolution of $\sim2$\arcsec\ towards a sample  of  146  high-mass star-forming IRAS regions in the range $-80\degr<l<40\degr$ and $|b|<2\degr$ \citep{Bro96,Fau04}. 
The overview paper (Paper\,I) presents  the source sample, the spectral setup, and the major goals of the survey.
Paper\,II addresses the relation between high-mass star formation and the dense gas distribution by investigating the star formation scaling relations inferred from different dense gas tracers 
(e.g., \hcop/\htcop , HCN/H$^{13}$CN).
Paper\,III includes the catalogues of candidate hot molecular cores and hyper/ultra compact HII regions, which provide an important foundation for future studies of the early stages of high-mass star formation across the Milky Way.

In the present paper, we make full use of the ATOMS data to gain insight  into the fragmentation and dynamical processes of the \filname\ cloud through observations of
its two massive, luminous  protostellar clumps, MM1 and MM2.
The paper is organized as follows: Section\,2 gives a brief description of the ALMA observations of the ATOMS survey, Section\,3 presents analysis of the ATOMS data, Section\,4 discusses the observed hierarchical fragmentation and the associated dynamical mass inflow/accretion scenario, and Section\,5   summarizes the results.

\section{ALMA Observations and data reduction}
\label{sec:observations}
Observations of the ATOMS survey consist of single pointings towards 146 IRAS clumps by both the Atacama Compact 7-m Array (ACA; Morita Array) and the 12-m array (C43-2 or C43-3 configurations) in Band\,3. 
Eight spectral windows (SPWs) were optimised to cover 11 commonly-used lines that includes the tracers of dense gas (e.g., HCO$^+$/H$^{13}$CO$^+$), hot molecular core gas (e.g., CH$_3$OH), shocked gas (e.g., SiO, and SO), and ionized gas (e.g., H${40}\alpha$). 
The basic parameters (e.g., rest frequency, transition) 
of these lines are listed in Table\,2 of Paper\,I. 
The SPWs\,1--6 are located at the lower sideband in the range [86.31, 99.40]\,GHz with spectral resolution
of $\sim 0.2-0.4$\,\vel\ for kinematic measurements, while SPWs\,7--8  in the upper sideband in  the range [99.46, 101.34]\,GHz each has a broad bandwidth of 1875\,MHz at a spectral resolution of $\sim 1.6$\,\vel\ for sensitive continuum measurements. 

The  data were calibrated in CASA 5.6 \citep{McM07}. 
We then imaged and cleaned the ACA and 12\,m-array data jointly using natural weighting (to optimise the signal-to-noise ratio) and taking $pblimit=0.2$, in the CASA {\it tclean} task,
for both continuum images and line cubes. 
Continuum images were created from line-free frequency ranges of SPWs\, 7--8 centred
at $\sim99.4$\,GHz while the spectral line cube of each SPW was produced with its native spectral resolution.
The resulting continuum image and line cubes for the 146 target clumps have angular resolutions $\sim 1\farcs2$--$1\farcs9$, 
and maximum recoverable angular scales $\sim 60\arcsec$.

In this paper, we analyze the I18507+0121 source from the ATOMS, whose ALMA observations cover the MM1 and MM2 clumps of G34. The analysis presented is carried out on the combined data sets unless specified otherwise.
The 12m+ACA combined continuum image of this source has a beam size of $1.9\arcsec \times 2.1\arcsec$, and 
a sensitivity of  $1 rms=0.3$\,mJy~beam$^{-1}$, which corresponds to a mass sensitivity of $\sim 0.1$\,\msun\ at the distance of the source for a dust temperature of $\sim 20$\,K (see Sect.\,\ref{subsec:result:cont} for the detailed mass calculation). 
For line cubes, only the
HCO$^+$~(1--0), H$^{13}$CO$^+$~(1--0), SiO~(2--1), SO~(3--2), CS~(2--1), and CH$_3$OH~2$_{\rm(1,1)}$--1$_{\rm(1,0)}$A lines 
are utilized for investigating the kinematics and dynamics of MM1 and MM2. 
The velocity resolutions for the HCO$^+$, H$^{13}$CO$^+$/SiO, and  SO/CS/CH$_3$OH lines are 0.1, 0.2, and 1.5\,\vel, respectively, and the sensitivity levels are $\sim 12$, $8$, and $3$\,mJy~beam$^{-1}$, respectively.

\begin{figure*}
\centering
\includegraphics[width=6.4 in]{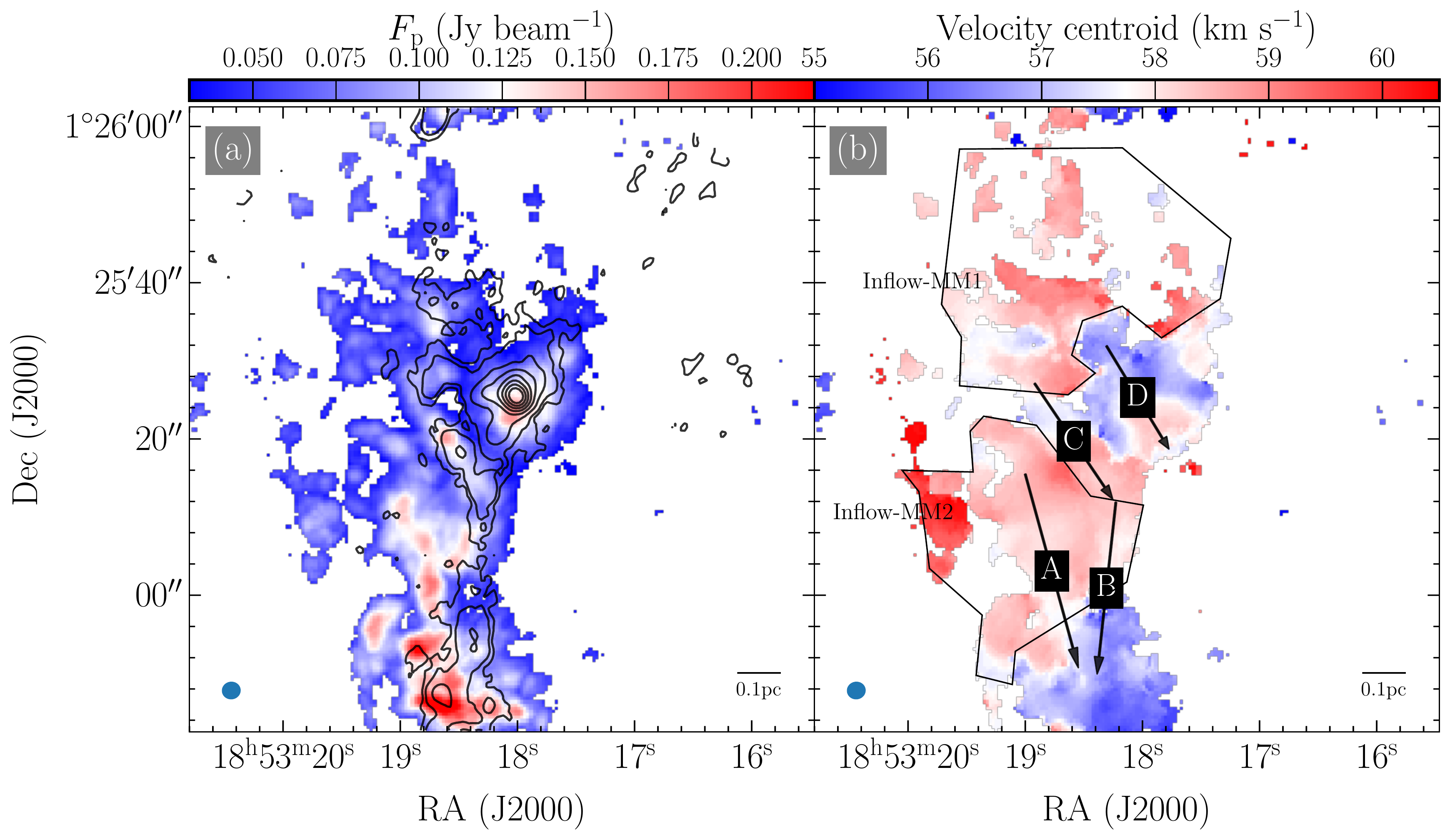}
\includegraphics[width=6.4 in]{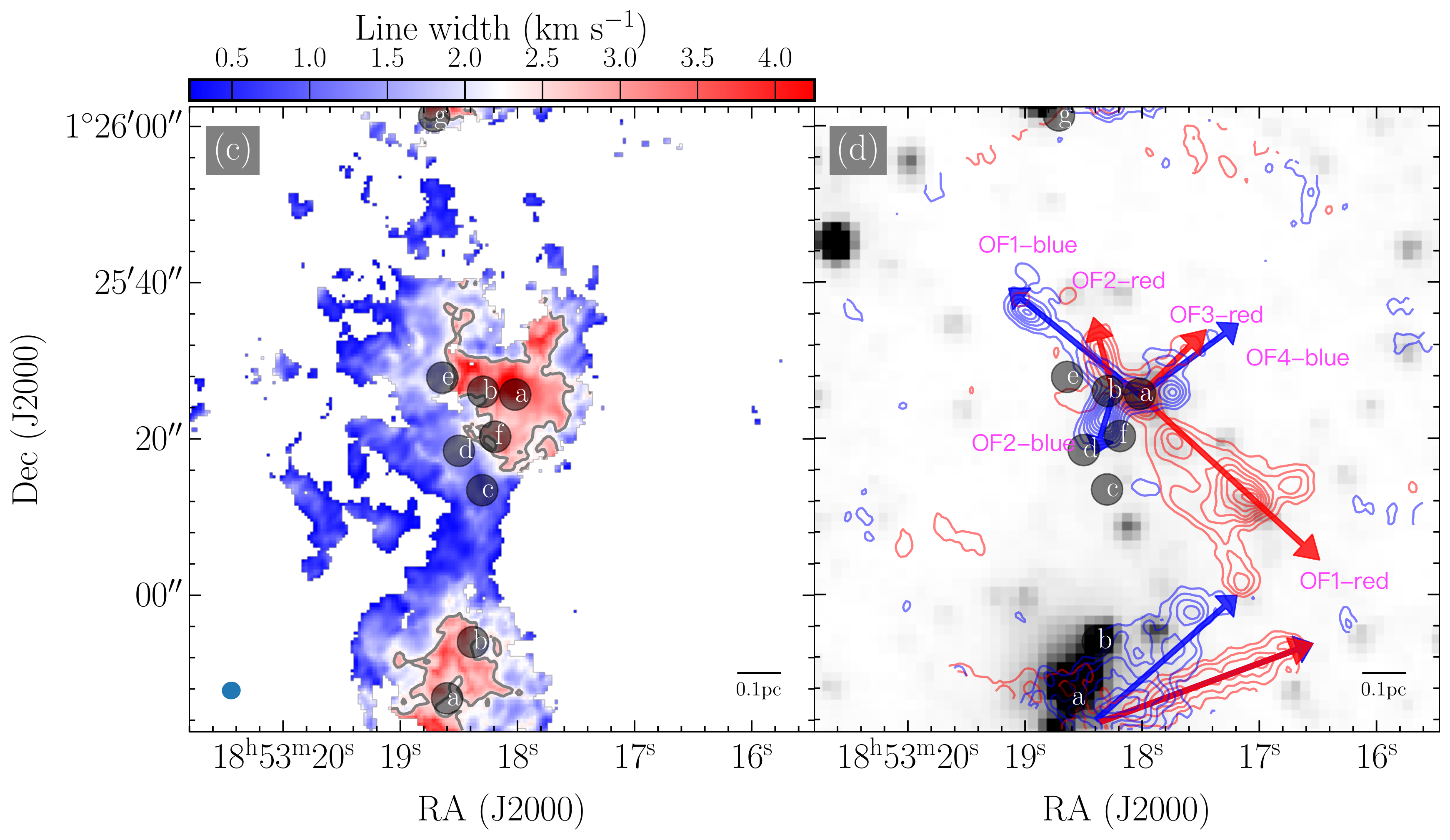}
\caption{ 
(a) Peak intensity map of \htcop~(1--0) for the MM1 and MM2 protostellar clumps in G34. 
 The 3~mm dust continuum emission is shown in black contours, starting at 3\,rms (rms~$\sim0.3$~mJy beam$^{-1}$) with the steps following the same power-law form as in Fig.\,\ref{fig:overview}. 
(b) Moment\,1 map of \htcop~(1--0). 
The arrows\,A--D mark the directions of the  observed velocity-coherent gradients.  
The polygons represent the gas inflow regions for each cluster of cores in MM1 and MM2. 
(c) Line width map of \htcop~(1--0).  
The contour represents a line width of $2.5$\,\vel. 
(d) CS~(2-1) outflows (contours) superimposed on the Spitzer 4.5\,$\mu$m image.  The red and blue arrows indicate the red and blue lobes of outflows, respectively. 
Labels\,a--g in panels\,(c) and (d) identify the dense cores in MM1 and MM2 protostellar clumps. 
In all panels, the map is displayed only at the positions where the peak intensity of the spectrum is $\geq5$ times the local noise level.
}
\label{fig:kin:maps}
\end{figure*}

\section{Results and analysis}
\label{sec:analysis}

\subsection{3\,mm continuum emission\label{subsec:result:cont}}
\begin{table*}
\centering
\caption{Continuum core parameters.}
\label{tab:cores}
\resizebox{18cm}{!}{
\begin{tabular}{ccccccccccccccc}
\hline\hline
  Name &            $RA$  &          $Dec.$ & $FWHM_{\rm dec}$  &              PA & $F_{\rm 3mm}^{int.}$ & $F_{\rm 3mm}^{p}$ &  $R_{\rm core}$ &   $T_{\rm gas}$  &  $M_{\rm core}$  & log($n_{\rm core}$) & $\Sigma_{\rm core}$ &   $V_{\rm lsr}$ &      $\Delta V$  & $\alpha_{\rm vir}$ \\
       &         (J2000)  &         (J2000) & ${\rm maj(\arcsec)  \times min(\arcsec)}$  &             deg &             mJy & mJy beam$^{-1}$ &    10$^{-2}$ pc &               K  &     $M_{\odot}$  &       cm$^{-3}$ &     g cm$^{-2}$ &     km s$^{-1}$ &     km s$^{-1}$  &                 \\

\hline
  MM1-a  &      18:53:18.02  &        1:25:25.7  &     2.3  $\times$     2.2  &    21.4  &   124.2  $\pm$     0.8  &    54.7  $\pm$     0.2  &     2.0  $\pm$     0.2  &   100.0  &     199  $\pm$      32  &     7.4  &    32.7  &    57.6  $\pm$     0.2  &     4.5  $\pm$     0.2  &     0.5  \\ 
  MM1-b  &      18:53:18.29  &        1:25:25.9  &     2.2  $\times$     2.0  &   102.1  &     9.6  $\pm$     0.1  &     9.6  $\pm$     0.2  &     1.9  $\pm$     0.2  &    27.6  &      58  $\pm$       9  &     7.3  &    11.0  &    57.1  $\pm$     0.2  &     2.9  $\pm$     0.2  &     0.6  \\ 
  MM1-c  &      18:53:18.30  &        1:25:13.6  &     3.5  $\times$     2.4  &    16.0  &     9.1  $\pm$     1.0  &     2.8  $\pm$     0.2  &     2.6  $\pm$     0.2  &    14.2  &     115  $\pm$      19  &     7.0  &    11.1  &    59.3  $\pm$     0.2  &     1.1  $\pm$     0.2  &     0.1  \\ 
  MM1-d  &      18:53:18.50  &        1:25:18.4  &     3.8  $\times$     2.5  &    32.0  &    10.3  $\pm$     1.0  &     3.0  $\pm$     0.2  &     2.7  $\pm$     0.2  &    13.8  &     136  $\pm$      22  &     7.0  &    12.1  &    58.6  $\pm$     0.2  &     1.8  $\pm$     0.2  &     0.1  \\ 
  MM1-e  &      18:53:18.64  &        1:25:28.0  &     4.5  $\times$     3.5  &   112.9  &     5.6  $\pm$     0.1  &     1.9  $\pm$     0.2  &     3.5  $\pm$     0.3  &    22.6  &      42  $\pm$       7  &     6.5  &     2.2  &    57.4  $\pm$     0.2  &     1.6  $\pm$     0.2  &     0.5  \\ 
  MM1-f  &      18:53:18.19  &        1:25:20.3  &     2.7  $\times$     2.1  &   152.5  &     6.9  $\pm$     0.1  &     5.5  $\pm$     0.2  &     2.1  $\pm$     0.2  &    18.6  &      65  $\pm$      11  &     7.2  &     9.5  &    56.9  $\pm$     0.2  &     3.3  $\pm$     0.2  &     0.8  \\ 
  MM1-g  &      18:53:18.71  &         1:26:1.3  &     3.7  $\times$     1.6  &   151.0  &    24.5  $\pm$     0.9  &     9.1  $\pm$     0.2  &     2.1  $\pm$     0.2  &    21.7  &     196  $\pm$      32  &     7.4  &    28.3  &    58.4  $\pm$     0.2  &     3.4  $\pm$     0.2  &     0.3  \\ 
  MM2-a  &      18:53:18.60  &        1:24:46.7  &     5.3  $\times$     4.0  &    23.2  &    62.3  $\pm$     1.5  &     9.8  $\pm$     0.2  &     4.1  $\pm$     0.3  &   100.0  &      97  $\pm$      16  &     6.4  &     3.8  &    56.5  $\pm$     0.2  &     2.5  $\pm$     0.2  &     0.7  \\ 
  MM2-b  &      18:53:18.38  &        1:24:54.1  &     6.7  $\times$     2.9  &     0.8  &    34.0  $\pm$     1.4  &     5.5  $\pm$     0.2  &     3.9  $\pm$     0.3  &    20.8  &     281  $\pm$      46  &     6.9  &    12.2  &    56.2  $\pm$     0.2  &     2.7  $\pm$     0.2  &     0.2  \\ 

\hline
\end{tabular}
}

\begin{flushleft}
{\bf Note:} $R_{\rm core}$ is derived from $R_{\rm eff}/3600\times \pi/180\times D$ given the relation $R_{\rm eff} = \sqrt{FWHM_{\rm dec}^{\rm maj}\times FWHM_{\rm dec}^{\rm min}}/2$, and the distance $D$ of the core.  Gas temperature for all the cores, except for MM1-a and MM2-a, is estimated from the kinetic temperature map obtained from VLA observations of ${\rm NH_{3}}$ by \citet{Lu 14}. For cores MM1-a and MM2-a, the temperature is assumed to be 100\,K given their association with a hot molecular core and an \uchii\ region, respectively. $V_{\rm lsr}$ and $\Delta V$ along with the associated errors are derived from single Gaussian component fit to the average spectrum of \htcop~(1--0) over each core. The errors for the fluxes result from the 2D Gaussian fitting in the core extraction, while the ones for $R_{\rm core}$,  $M_{\rm core}$, and $n_{\rm core}$ are mainly due to the distance uncertainty.
\end{flushleft}
\end{table*}

Figure\,\ref{fig:overview} illustrates the overall morphology and location of the MM1 and MM2  protostellar clumps of \filname\ in the 
{\it Spitzer} 8.0\,$\mu$m image overlaid with ATLASGAL 870\,\um\ dust continuum in panel\,(a), and 
with the ATOMS 3\,mm continuum data in panel\,(b).  
 The ATOMS continuum image presented in this figure and Fig. \ref{fig:kin:maps} is not corrected for the primary beam response. This is done only for the purpose of display since it enables a uniform noise level to be shown in the map. But for the analysis of the core properties, the primary-beam corrected 3\,mm image is used (see Paper\,III for more details).
The dashed circle in Fig.\,\ref{fig:overview}\,a marks the field of view (80\arcsec\ in diameter) of the resulting combined image of ATOMS  used in this study.
As seen, the MM1 clump is fully covered while only half of the MM2 area is observed by the ATOMS.
The large-scale filamentary cloud overall appears dark against the background emission at 8\,\um.  Also marked in the figure is the chain of the three millimeter clumps (i.e., MM1, MM2, and MM4), identified by \citealt{Rat06} from 1.2\,mm continuum  observations. The orientation of the clumps along the filament is seen to be replicated
on smaller scales as well, as is evident from the north-south spread of the detected cores in the 3\,mm continuum of the ATOMS data. The 3\,mm continuum emission observed across the entire region mainly comes from thermal dust emission since ATOMS detected no  \halpha\ emission.  Further, only two very compact centimetre sources of size $\sim 1$\,\arcsec\  are detected by \citet{Ros16}, which are confined within the centres of the MM1 and MM2 clumps.
 
The high angular resolution of the ATOMS 3\,mm  continuum map allows to identify the dense cores where stars form. 
Paper\,III adopted a combination of the {\it Dendrogram} algorithm and CASA-{\it imfit} function to extract cores.  
As discussed by these authors, the former technique does not always provide good measurements of the core parameters on size and position angle, while the latter performs better in this regard through a two-dimensional Gaussian fit to the emission. 
Following this  approach, nine cores are extracted from  the 3\,mm continuum  map with seven (i.e., MM1a--g, see Fig.\,\ref{fig:overview}b)  in MM1 and two  (i.e., MM2a--b)  belonging to MM2. 
Note that the MM1-b core was 
manually located and then extracted with CASA-{\it imfit} since it was not automatically  detected by {\it Dendrogram} due to its close proximity and small intensity contrast relative to the neighbouring MM1-a core. The additional extraction of MM1-b was  driven by the associated outflows (see Fig.\,\ref{fig:kin:maps}d). Its existence
was further verified through a careful visual examination of the radial intensity profile along the 
direction connecting both MM1-b and MM1-a sources, where the weak and 
strong intensity peak components correspond to the two sources, respectively.

Overall, the result of core extraction in this work  matches that of Paper\,III with the exception of the two cores, MM1-g, and MM2-a.
Both of them are located outside a circle of radius 36\arcsec\ around the image centre, which was imposed as a mask for batch  extraction of cores from all of the 146 ATOMS clumps, and thus ignored in Paper\,III. 
Our ATOMS observations have revealed the  most complete population of cores (i.e., a cluster of seven cores) in MM1, compared with the previous SMA observations \egcite{Rat11}, in which
only the  MM1-a core was revealed.  This could  be a consequence of either the  missing-flux effect due to the limited {\it uv} coverage by SMA or the poor sensitivity.
Additionally, due to the incomplete coverage by the ATOMS for MM2, we believe that the population of cores in this clump has not been fully unveiled, and  this clump could also be harbouring a cluster of cores. 

The measured parameters of the nine cores  are listed in Table\,\ref{tab:cores}, including the deconvolved sizes of the major and minor axes  (Col.\,4), the position angle (Col.\,5), the core-integrated 3\,mm flux (Col.\,6), and the core-peak flux (Col.\,7). 
Note that both MM1-a and MM2-a cores have an associated compact centimeter source, each with a size of $\sim 1$\,\arcsec. 
The two sources coincide with 
an early B-type star in MM1 (see Fig.\,\ref{fig:overview}, \citealt{She07}), and an \uchii\ region in MM2. 
In addition, the centimeter sources in MM1-a and MM2-a correspond to partially optically-thick free-free emission and non-thermal emission, respectively, with the former having a spectral index of $\alpha=0.7$ in the frequency range 4.9\,GHz to 
25.5\,GHz and the latter having $\alpha=-0.5$ \citep{Ros16}. 
Given a total flux density of $\sim0.4$\,mJy and $9.5$\,mJy at 4.9\,GHz for the centimeter sources in MM1-a, and MM2-a, respectively, the relation $S_{\nu}\propto\nu^\alpha$ yields a flux density of $\sim 3$\,mJy and $2.1$\,mJy, respectively, at 3\,mm, which is negligible compared to the total fluxes of the two cores.

The mass, $M_{\rm core}$, and the column density,  $N_{\rm core}$, of the cores were calculated following Eqs.\,B1--B2 of Paper\,III. 
In the calculation,  the gas temperature for all the cores, except for MM1-a and MM2-a, is estimated from the kinetic temperature map obtained from the VLA $3$\arcsec-resolution observations of the ${\rm NH_{3}}$~(1, 1) and (2, 2) inversion lines by \citet{Lu 14}. For MM1-a and MM2-a, the temperature is assumed to be 100\,K due to the presence of an associated hot molecular core, and an \uchii\ region (\citealt{Rat11}, Paper\,III), respectively. 
In addition, a gas-to-dust mass ratio of $R_{\rm gd}=100$ was adopted. Recent
works propose higher values ($R_{\rm gd}^{\prime}$) like 150 \citep{Dra11} or 162 \citep{Pet17}. One can estimate $M_{\rm core}$ and
 $N_{\rm core}$ with different gas-to-dust mass ratios simply by multiplying a scaling factor of $R_{\rm gd}^{\prime}/R_{\rm gd}$. In the above calculation the core flux includes background emission,  
 which could lead to the overestimation of the core masses. In practice, the background emission is difficult to be accurately subtracted especially from the high-resolution ATOMS data that has already filtered out a significant portion of large-scale components. For a conservative estimate, we assume a  constant value of 1\,rms level for the background emission. Using this, it is seen that the median value of the new core masses decreases by $\sim 25\%$ where the decrease in mass is mostly seen in the relatively low-density cores while the more massive and dense ones remain more or less unchanged.

Furthermore, the mass surface density   can be derived  from $\Sigma_{\rm core} = M_{\rm core}/(\pi R_{\rm core}^2)$, while the number density 
from $n_{\rm core} = N_{\rm core}/2R_{\rm core}$,  
where $R_{\rm core}$ is the core radius equal to the geometric mean of $FWHM_{\rm maj}^{\rm dec}$ and $FWHM_{\rm min}^{\rm dec}$ at the core distance.
The above derived parameters can be found in Cols.\,9--11 of Table\,\ref{tab:cores}. 
In summary, we find 
$R_{\rm core}\sim$~0.02--0.04\,pc with a median value of 0.03\,pc, $M_{\rm core}\sim$~42--281\,\msun\ with a median value of 115\,\msun, 
$n_{\rm core}\sim$0.2$\times10^{7}$--2.8$\times10^{7}$\,\pcmcu\ with a median value of $1.1\times10^{7}$\,\pcmcu,
and $\Sigma_{core}\sim$~2--32\,g\,\pcmsq\ with a median value of 11\,g~\pcmsq. 
If a theoretical threshold of 1\,g~\pcmsq\, above which the cores most likely form high-mass stars, is assumed \citep{Kru08}, then the entire population of cores detected in this study have the ability to form high-mass stars.

 Following the method described above, we also estimated the mass, $M_{\rm cl}$, and the number density, $n_{\rm cl}$, of the \filname\ cloud (80\arcsec\ in size) considering the region within the field of view of the ATOMS observations. Since the kinetic temperature map of \citet{Lu 14} is obtained using interferometric observations, structures with spatial scales larger than that of the ATOMS cores are resolved out. Hence, for these calculations, we assume a cloud-average dust temperature of $\sim 21$\,K, which was measured over the field of view of the ATOMS observations from the publicly available dust temperature map created using the  point processing mapping (PPMAP) technique, a 
state-of-the-art spectral energy distribution fit method \citep{Mar17}. 
The assumption of this temperature is in good agreement with the cold nature of the IRDC cloud \egcite{Car98,Car20,Rat06, Soa19}. Taking the integrated flux of 480\,Jy at 870\,\um\ from the ATLASGAL image of the cloud leads to $M_{\rm cl}\sim4.2\times10^4$\,\msun, and $n_{\rm cl}\sim2.6\times10^5$\,\pcmcu. 
The mass conversion  efficiency from cloud to cores is rather low  at about $3\%$, although the masses of the cores are far above the mass sensitivity (i.e., $\sim 0.3$\,\msun\ for a 3\,rms detection) of the observations, implying that most of the cloud gas is not efficiently transformed into cores.

\subsection{Molecular gas emission \label{subsec:result:molecular}}
\begin{figure}
\centering
\includegraphics[width=3.4 in]{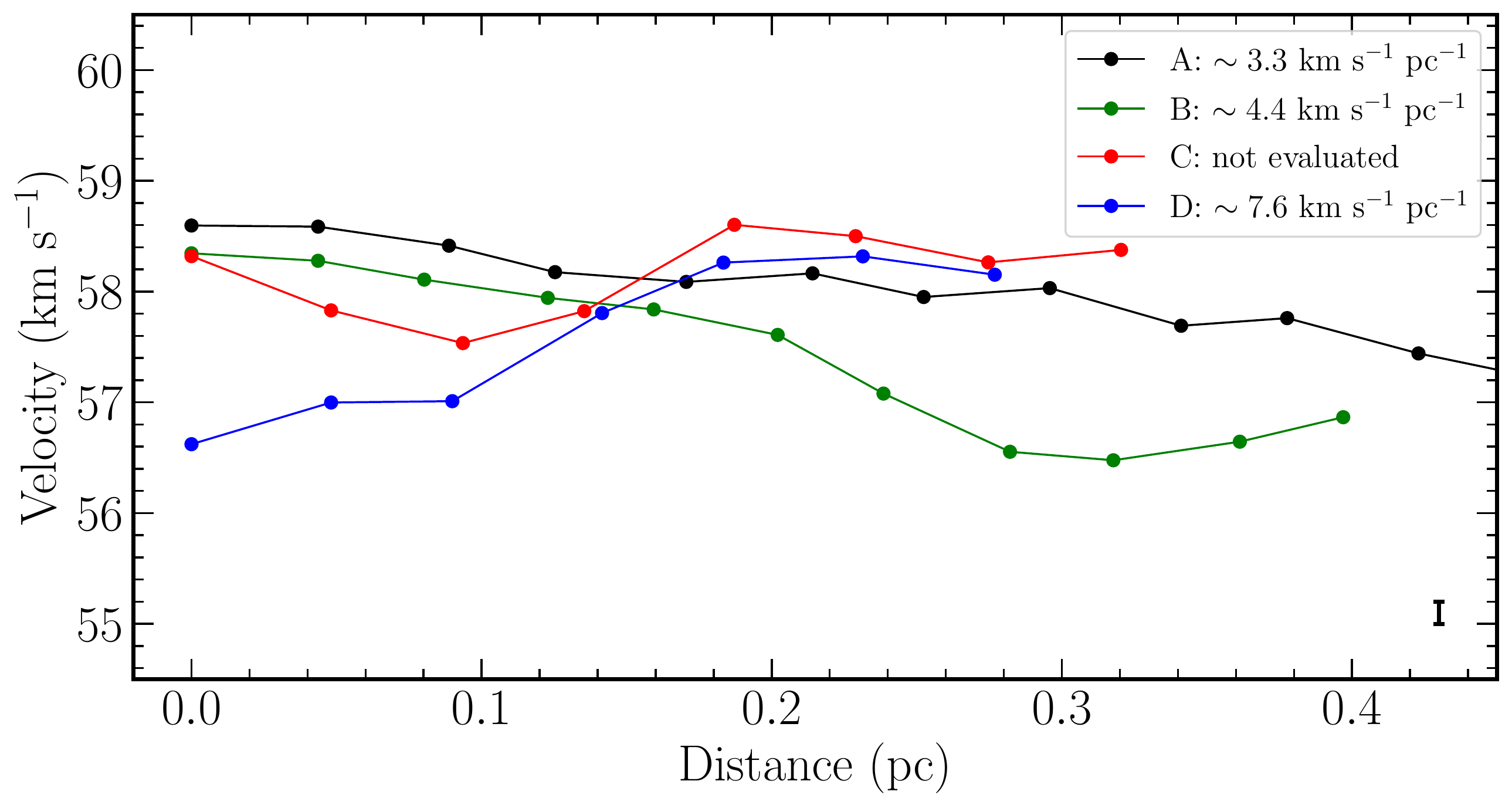}
\caption{ Velocity distribution derived from \htcop~(1--0) along the four selected directions A--D, which are indicated in Fig.\,\ref{fig:kin:maps}b. The typical error bar of the velocity gradient is shown at the bottom right.
}
\label{fig:vel:grad}
\end{figure}

\begin{figure}
\centering
\includegraphics[width=3.4 in]{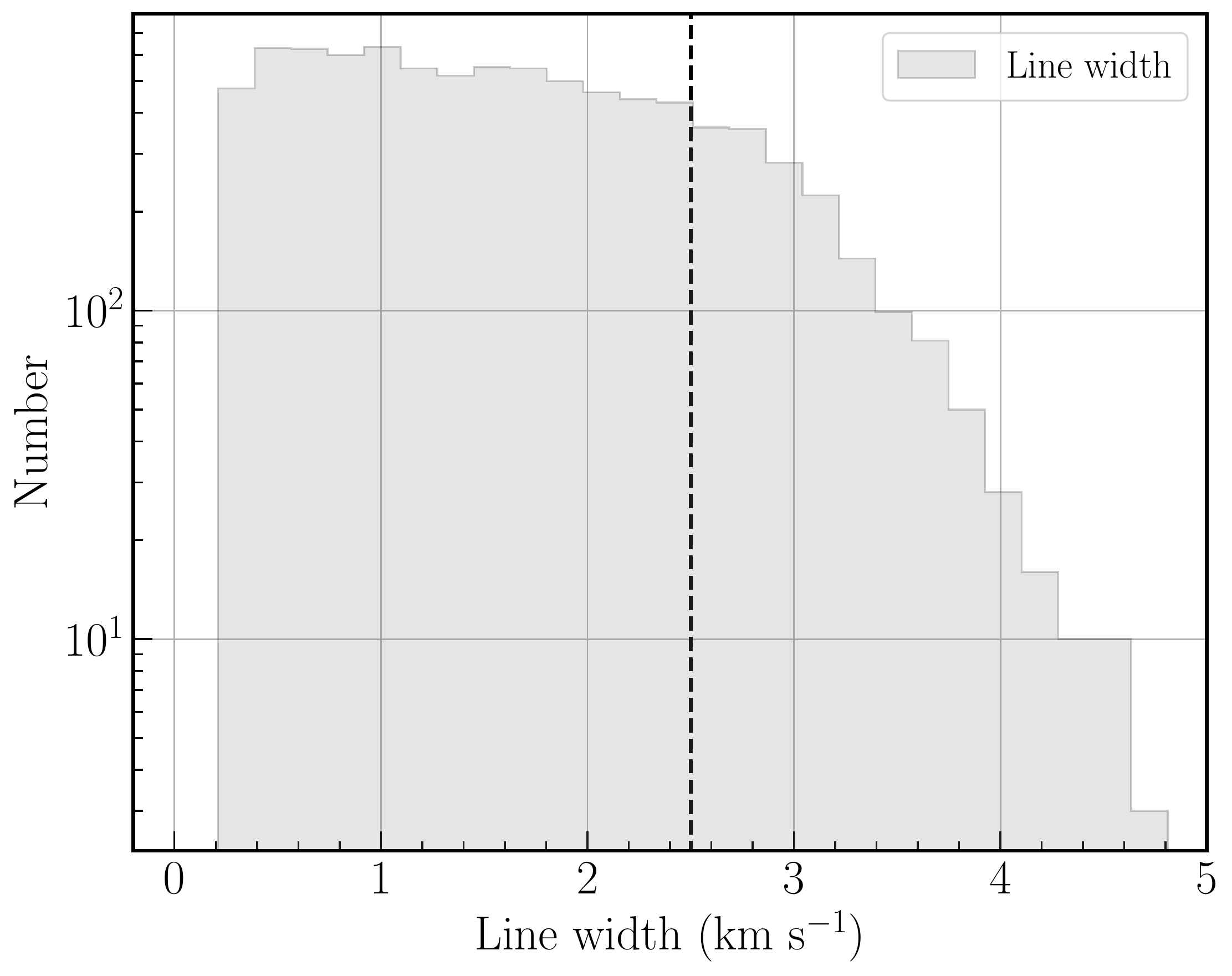}
\caption{Distribution of the line widths derived from \htcop~(1--0). The y-axis shows the number of pixels. An inflection in the distribution can be found at $\sim 2.5$\,\vel (indicated by the dashed vertical line).
}
\label{fig:kin:linewidth:hist}
\end{figure}

The average spectrum of \htcop~(1--0) over the entire region investigated here reveals a systemic velocity of $V_{\rm lsr}=57.6$\,\vel.
In general, \htcop\ emission is treated as relatively optically thin. However, this may not be the case for the densest part of cores. Following Eq.\,1 of \citet{Liu20a}, we calculate the optical thickness from the peak intensity of \htcop\ of the nine cores with the assumptions of local thermodynamic equilibrium (LTE).
This yields optical depth values in the range of $0.04-0.39$ and hence supports the optically thin assumption for the \htcop\ emission.
Besides, extended emission of \htcop\ appears as a single-peaked spectral profile across the entire region at a 
high spectral resolution 0.2\,\vel, which makes the \htcop~(1-0) line a good probe for the spatial distribution of molecular gas within the MM1 and MM2 clumps.

Figure\,\ref{fig:kin:maps}a displays the map of the peak intensity of \htcop~(1--0). 
For comparison, the 3\,mm dust continuum is also overlaid as contours in the figure.  \htcop~(1-0) gas and 3\,mm dust
continuum distribution are found to match each other well in dense regions, with the former being much more extended than the latter across the entire region.
In particular, a bright branch of \htcop\ emission, which
appears to stretch towards the cluster of cores (MM2a--b) in MM2, does not have detectable continuum emission at the current sensitivity.
The nondetection of  dust continuum indicates that this branch of \htcop\ emission could be  ambient gas having lower column density than that concentrated in the clusters of cores with apparent dust emission.

Figure\,\ref{fig:kin:maps}b presents  the moment\,1 map of \htcop~(1--0) that can reveal the global velocity field. 
Velocity gradients can be seen towards the cluster of cores in both MM1 and MM2.
In particular, the velocity gradients towards the cluster of cores in MM2 appear to match the  bright branch of \htcop\ gas emission mentioned above,
suggesting that ambient gas is being accelerated by the gravity of the cluster of cores in MM2, and thus inflowing towards them (see Sect.\,\ref{subsec:disc:dynamic} for additional discussion).
Quantitatively, the velocity gradients are evaluated along the four directions (indicated by arrows A--D in Fig.\,\ref{fig:kin:maps}b). These four directions are visually identified after a careful examination of the moment 1 map.
We find from Fig.\,\ref{fig:vel:grad} that the gradients lie in the range of $\sim 3$--$8$\,\vel~pc$^{-1}$. Of particular interest is the gradient along the direction D, 
which appears to be the signature of rotational motion (see Sect.\,\ref{subsec:result:rotate} for more analysis).

Figure\,\ref{fig:kin:maps}c shows the line width map of \htcop~(1--0) reflecting the global kinematics of the entire region investigated here. The line width tends to be enhanced around the centres of the clusters of cores in both MM1 and MM2. Given that the \htcop~(1--0) emission is seen to be optically thin, the effect of optical depth can be ruled out and this enhancement can be attributed to intense star-forming feedback (e.g., energetic outflows and stellar winds) due to the presence of the luminous YSO in MM1 and the \uchii\ region in MM2 \citep{She07,Liu20a}. Moreover, the dynamical motions of the gravity/turbulence-driven inflows toward the centres of MM1 and MM2  can be an additional source for the enhanced line width.

To quantitatively describe this enhancement, we plot, in Fig.\,\ref{fig:kin:linewidth:hist}, the distribution of the line widths for the entire region.
A distinct inflection at the line width of $\sim2.5$\,\vel\ is evident on visual inspection of the distribution. 
This turnover could represent the threshold above which the line
width is enhanced by strong feedback from star formation and probably by the dynamical gas inflows, and below which the gas kinematics are less affected. 
This threshold can also be found in 
Fig.\,\ref{fig:kin:maps}c where the threshold of $\sim2.5$\,\vel\ (in gray contour) does separate the weak- and strong-feedback areas well.
The mean line width is $\sim 2.9$\,\vel\ in the area of strong feedback, and $\sim 1.3$\,\vel\ elsewhere.

\subsection{Molecular outflows\label{subsec:result:outflows}}
\begin{table*}
\centering
\caption{CS~(2-1) outflow parameters.}
\label{tab:outflow_cs}
\resizebox{18cm}{!}{
\begin{tabular}{cccccccccc}
\hline\hline
Outflow & Lobes  &           $<L>$ &            $<V>$  &   $M_{\rm out}$ &   $P_{\rm out}$ &   $E_{\rm out}$ &   $F_{\rm out}$ &   $t_{\rm dyn}$  & $\dot M_{\rm acc}$  \\
      &       &              pc &     km s$^{-1}$  &     $M_{\odot}$ & $10^{2}M_{\odot}$ km s$^{-1}$ &    $10^{46}$erg & $10^{-3}M_{\odot}$ km s$^{-1}$ yr$^{-1}$ &        $10^3$yr  & $10^{-5}M_{\odot}$ yr$^{-1}$  \\

\hline
 MM1-OF1  &    B+R  &    0.47  $\pm$    0.07  &    14.6  $\pm$     1.5  &    34.9  $\pm$     1.7  &     5.1  $\pm$     0.3  &     7.8  $\pm$     0.4  &    16.2  $\pm$     5.0  &     3.1  $\pm$     0.8  &     3.0  --     6.0  \\ 
 MM1-OF2  &    B+R  &    0.18  $\pm$    0.03  &    14.9  $\pm$     1.5  &     8.0  $\pm$     0.4  &     1.2  $\pm$     0.1  &     1.8  $\pm$     0.1  &    10.1  $\pm$     3.3  &     1.2  $\pm$     0.3  &     2.0  --     4.0  \\ 
 MM1-OF3  &      R  &    0.23  $\pm$    0.02  &    18.2  $\pm$     1.8  &     1.8  $\pm$     0.1  &     0.3  $\pm$     0.1  &     0.6  $\pm$     0.1  &     2.6  $\pm$     0.6  &     1.3  $\pm$     0.2  &     0.4  --     0.7  \\ 
 MM1-OF4  &      B  &    0.14  $\pm$    0.01  &    12.0  $\pm$     1.2  &     1.1  $\pm$     0.1  &     0.1  $\pm$     0.1  &     0.2  $\pm$     0.1  &     1.1  $\pm$     0.3  &     1.2  $\pm$     0.2  &     0.2  --     0.3  \\ 
\hline 
   Total  &    --     &         --                &           --              &    44.7  $\pm$     1.7  &     6.6  $\pm$     0.3  &    10.2  $\pm$     0.4  &    29.0  $\pm$     5.0  &           --              &     5.4  --    10.7  \\ 

\hline
\end{tabular}
}

\begin{flushleft}
Note: B and R in Col.\,2 stand for the blue and red lobe of outflows, respectively. 
\end{flushleft}
\end{table*}
 Previous single-dish as well as interferometric observations have shown the presence of outflows in the MM1 and MM2 clumps \egcite{Rat11,Zha14,Liu20a}.
With the new observations from the ATOMS survey that include several commonly-used outflow tracers like \hcop, CS, SiO, SO, and \chthoh, we
identify, in different tracers, the outflows associated with the two clusters of cores in MM1 and MM2. 
In the identification, the outflowing gas velocities of each tracer (line) are taken from the line wings of its spectrum, which lie outside the full
width of half maximum of the spectrum. The spectrum used is averaged over the entire region investigated here.
The blue- and red-shifted outflowing gas emission,  in sequence, is integrated  over  the velocity ranges (1) [44.0, 55.0]\,\vel\ and [60.0, 76.0]\,\vel\ for \hcop~(1-0),
(2) [40.0, 51.4]\,\vel\ and [63.4, 82.0]\,\vel\ for CS, 
(3)  [42.0, 51.3]\,\vel\ and [63.5, 74.0]\,\vel\ for SiO, (4) [42.0, 52.7]\,\vel\ and [62.2, 78.0]\,\vel\ for SO, 
and (5) [46.0, 54.2]\,\vel\ and [60.6, 66.0]\,\vel\ for \chthoh.

In total, six outflows are identified with four within the MM1 clump (i.e., MM1-OF1 to MM1-OF4) and two within MM2. 
All of the outflows are identifiable in the five tracers with the exception of MM1-OF3, which is visible only in 
CS and SiO emission. The outflows identified with each tracer are plotted in Figs.\,\ref{fig:kin:maps}d and \ref{fig:outflow:gallary}. For comparison, the {\it Spitzer} 4.5\,\um\ is overlaid, which is thought to be an indicator of shocked gas.
 As seen, the outflow gas extent does not correspond to the extended 4.5\,\um\ emission well. From the figure, 
MM1-OF1/OF3/OF4 appear to be driven by the embedded source(s) in MM1-a while MM1-OF2 by the source(s) in MM1-b. 
In MM1, all of the outflows except for MM1-OF1 are distinct from the results of \citet{She07}. 
The low-resolution (angular resolution of 3\farcs6) CO~(1-0) observations presented by these authors do not reveal the presence of MM1-OF2 nor do they distinguish between the MM1-OF3 and MM1-OF4 outflows.
Although both MM1-OF3 and MM1-OF4 outflows display single lobes, these have a high possibility of being associated with the MM1-a core due to the outflowing velocities ranging around the systemic velocity of MM1-a.

In a later study by \citet{Ise21}, based on higher 0\farcs8-resolution observations of CO~(2--1), the two lobes of MM1-OF2 were reported as separate outflows but not as a bipolar outflow. 
 Since the driving source(s) of the outflows within MM2 (i.e., MM2-OF1/OF2) could not be probed due to the incomplete coverage of ATOMS, 
they are not considered for further analysis.
In addition, we find that the outflows identified with the new observations are more collimated than those seen from the CO~(1--0) observations by \citet{She07}, suggesting that the outflow tracers used here, that have much higher critical density than CO~(1--0), could be probing the compact, highly-collimated jets.  

\subsection{Outflow parameters in MM1\label{subsec:result:outflowsMM1}}
Parameters such as the momentum, dynamical age, and ejection rate are valuable for characterizing the outflows.
In principle, both CS and SiO emission can be used for the calculation since  from Figs.\,\ref{fig:kin:maps}d and \ref{fig:outflow:gallary} they both are found to trace the extent of all of the outflows better than the other three tracers. 
Since the shock-sensitive SiO is preferentially enhanced in shocked regions, and less abundant elsewhere, the core-scale abundance 
is required to more properly estimate the abundance-related parameters like the momentum. 
Since no core-scale measurement of SiO abundance is available for the cluster of cores in MM1, we consider CS emission only,
which can be abundant in non-shocked as well as shocked regions, and  in general is not as sensitive to shocked gas as SiO emission.

Following \citet{She07}, an inclination angle of 45\,\degr\ was assumed for simplicity, which minimizes the errors introduced by inclination effects for outflows with unknown orientation.  
The total outflowing gas mass, $M_{\rm out}$, was inferred from $\Sigma M_{i}$, where the gas mass, $M_{i}$, in the velocity channel $i$  was calculated from Eq.\,4 of \citet{Liu20a} assuming optically-thin emission, LTE conditions,
and a CS abundance of $5.8\times10^{-10}$ inferred by \citet{Liu20a}.
In addition, the clump-average dust temperature of 38\,K  as quoted in \citet{Liu20a} was taken as the excitation temperature in the calculation.
The momentum, $P_{\rm out}$, was derived from $\Sigma M_{i} v_{i}$ and the kinetic energy, $E_{\rm out}$, from $\frac{1}{2} \Sigma M_{i} v_{i}^2$ with 
$v_{i}$ defined as the central velocity of the $i^{\rm th}$ channel relative to $V_{\rm lsr}$. 
The dynamical age, $t_{\rm dyn}$, was estimated from $<L>/<V>$, where
$<L>$ is the average length of the red and/or blue lobes of outflows and $<V>$ is the intensity-weighted mean velocity defined as $P_{\rm out}/\Sigma M_{i}$. 
Accordingly, the outflowing mass rate, $\dot M_{\rm out}$,
is given by $\Sigma M_{i}/t_{\rm dyn}$, and the mechanical force, $F_{\rm out}$, by $P_{\rm out}/t_{\rm dyn}$. 

Table\,\ref{tab:outflow_cs}  summarises the above-derived parameters for the CS outflows within MM1.
The parameter errors  mainly arise from the uncertainties of the distance and/or velocity measurements. 
The global properties of all of the outflows in MM1
(e.g., $M_{\rm out}$, and $P_{\rm out}$) agree with those derived from CO~(1--0) interferometric observations by \citet{She07}, who treated all outflows in MM1 as a  single outflow.
Further, the estimated dynamical ages of all outflows suggest that MM1-OF1 is the oldest while the lifetimes of MM1-OF2/OF3/OF4 are comparable, {whereas all of the outflows are very young with respect to the average dynamical time of order of 10$^{4}$~yr over near 400 molecular outflows catalogued by \citet{Wu 04}.}

If we assume momentum-driven outflows in protostellar-jet/outflow systems \egcite{Mas93, God20} along with momentum conservation, the relation $\dot M_{\rm jet} V_{\rm jet} = \dot M_{\rm out} V_{\rm out} = F_{\rm out}$ follows, where $\dot M_{\rm jet}$ and  $V_{\rm jet}$  are the mass-loss rate and the speed of the jet, respectively. 
The jet speed for massive outflows was observationally measured to be around 500\,\vel\ in several proper motion studies of radio continuum jets \egcite{Mar95}.
In protostellar jet/outflow systems, the ejection rate is expected to correlate with the accretion rate via the relation 
$\dot M_{\rm acc} = (1+f_{\rm jet})/f_{\rm jet} \times \dot M_{\rm jet}$. 
The fraction $f_{\rm jet}$, defined as the ratio between the accretion mass rate and the mass-loss rate through the jet, is poorly constrained by observations, but predicted in models to be 0.2--0.5 \egcite{Off14,Kui16}. 
This $f_{\rm jet}$ range yields a range of  the mass accretion rates for each outflow (last column of Table\,\ref{tab:outflow_cs}). 
If all of the outflows in MM1 are treated as  a single entity, the total mass accretion rate onto the protostars will be in the range [5, 11]$\times10^{-5}$\,\msun\,yr$^{-1}$,  which is in agreement with rates estimated for other high-mass star-forming systems \egcite{Zha05,Zha13,Yua18}

\subsection{A rotating envelope within MM1? \label{subsec:result:rotate}}
\begin{figure}
\centering
\includegraphics[width=3.4 in]{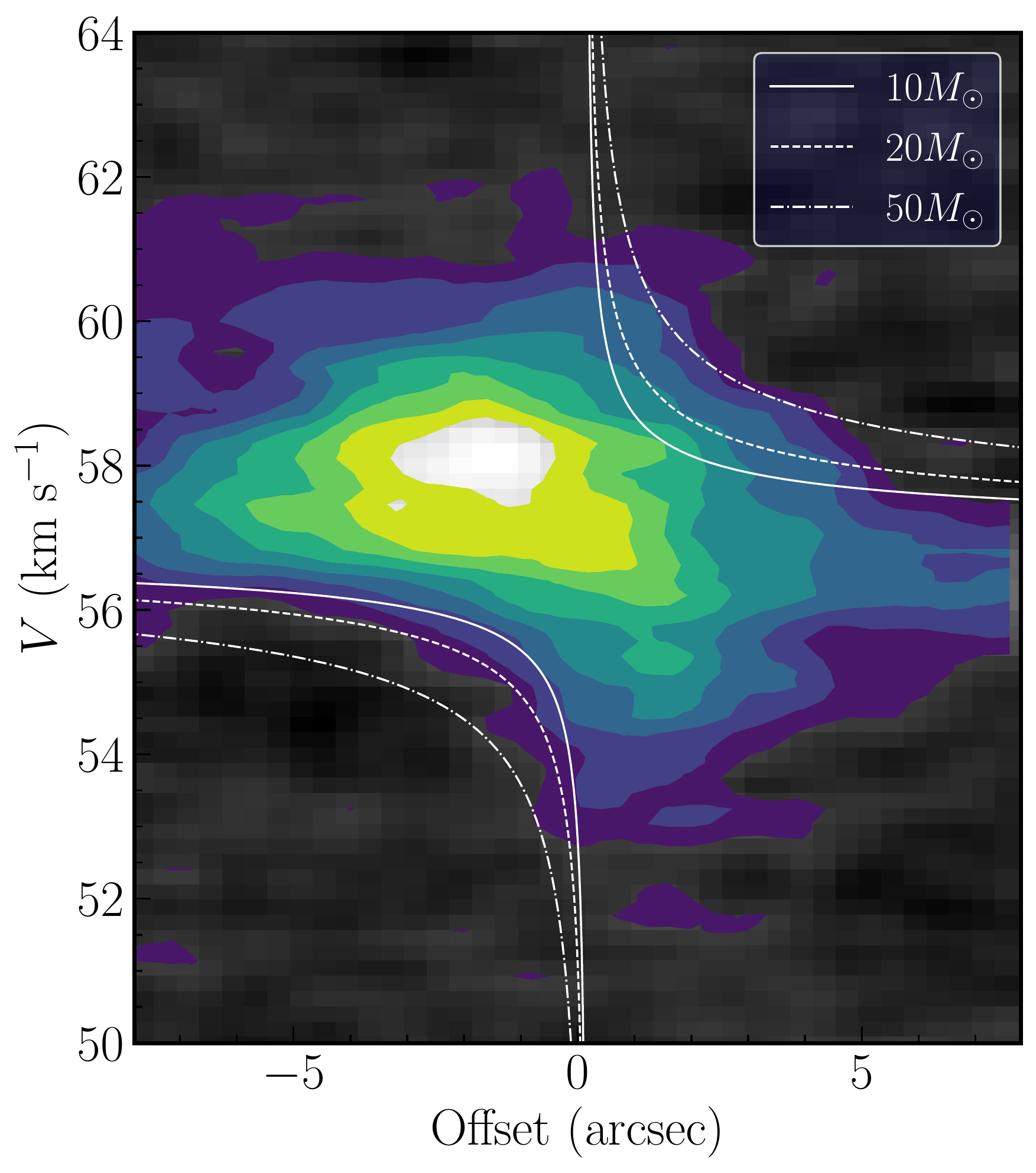}
\caption{ Position velocity diagram of \htcop~(1-0) for \filname--MM1.
The position offset is along the D direction indicated in Fig.\,\ref{fig:kin:maps}b. Overlaid on the image are 
models for Keplerian rotation with central masses of 10, 20, and 50\,\msun. 
}
\label{fig:MM1:pvd}
\end{figure}
The extended \htcop\ emission presents a clear southwest-northeast (SW-NE) velocity gradient (i.e., along the direction D in Fig.\,\ref{fig:kin:maps}b), with red-shifted emission to the southwest and blue-shifted emission to the northeast of the
MM1-a core. 
This observed velocity gradient has already been reported in higher 0\farcs6-resolution observations of \chthoh\ at $\sim 345$\,GHz with SMA by 
\citet{Rat11} (see right panel of their Fig.\,6). 
Ruling out the possibility of the influence from the large-scale MM1-OF1 outflow, these authors interpreted the velocity gradient as arising from a rotating structure surrounding the protostar(s) embedded in the core. 
Note that the SW-NE velocity gradient shown here is seen only in emission of \htcop\ but not in other species including the complex organic molecules (COMs), such as \chthoh, which are hot-gas tracers. This could perhaps be attributed to either the different transitions  involved or the different spatial resolutions  of the two data sets. For example, the transition responsible for \chthoh\ emission at $\sim90$\,GHz in the ATOMS data has a much lower upper energy temperature ($\sim20$\,K) than that of the transition of \chthoh\ at $\sim345$\,GHz  where the temperature is $>100$\,K. 
The \chthoh\ emission at $\sim90$\,GHz could therefore be tracing different physical regions than those by \chthoh\ at $\sim345$\,GHz. 
 Besides, the resolution of the ATOMS $\sim 2$\arcsec\ (corresponding to $\sim 0.035$\,pc at the distance of the \filname\ cloud) is around 3.3 times lower than that of the SMA observations of \citet{Rat11},  and hence, the \chthoh\ emission of the ATOMS may not be able to resolve the velocity structure at the small-scale, local area of hot gas of around 0.004\,pc as revealed in \citet{Rat11}.

To better understand the velocity distribution along the direction of the velocity gradient (i.e., direction D in Fig.\,\ref{fig:kin:maps}b), we plot the position-velocity diagram in Fig.\,\ref{fig:MM1:pvd}.
It reveals a typical butterfly-shaped appearance that is characteristic of Keplerian rotation pattern, which might be another indication of the existence of a rotating envelope.
Quantitatively, the velocity pattern can be fitted with a Keplerian rotation curve around an $Msin^2\theta \sim$10--50\,\msun\ protostar, where $\theta$ is defined as the inclination angle
between the rotation axis and the line of sight. Although this range of masses is consistent with 
the luminosity ($10^{4.3}$\,\lsun) and spectral type (B0) determined from the  SED of the MM1-a core by \citet{Rat06}, it is still 
rather small compared with the core's mass ($\sim 200$\,\msun, see Table\,\ref{tab:cores}) and is
inconsistent with the assumption of the Keplerian rotation that requires the gas mass to be negligible with respect to the central mass \egcite{Ces19}. 
This inconsistency can be alleviated if the rotating envelope is sufficiently inclined. 
For example, if $Msin^2 \theta \sim$10\,\msun, the stellar mass can exceed the core mass of 200\,\msun\  for $\theta<13$\degr.

The mass discrepancy can also be addressed if one considers the mass of $\sim 8$\,\msun\ of the more concentrated $\sim 0.02$\,pc diameter core, as defined by \citet{Rat11}, rather than the large mass of the MM1-a core $\sim 0.04$\,pc in diameter estimated with the ATOMS data. 
Even if this is the case, the rotating structure is too large, around 10\arcsec (0.18\,pc), to be arising from an accretion disk, which is generally observed to have a size of a few hundred AUs \egcite{Rat11,Mos21}. 
We therefore suggest that the large-scale velocity-coherent gradient revealed in \htcop\ emission could be ascribed to a relatively large, rotating structure.

\subsection{Virial Analysis \label{subsec:result:virial}}

\begin{table}
\centering
\caption{Jeans parameters on cloud and clump scales. }
\label{tab:fragment}
\resizebox{6cm}{!}{
\begin{tabular}{lcc}
\hline\hline
Parameter & Cloud & Clump \\
\hline
$L_{\rm Jeans}^{\rm th}$ (pc):     &    $\sim 0.1$  &  $\sim 0.06$ \\
$M_{\rm Jeans}^{\rm th}$ (\msun):  &    $\sim 2.6$  &  $\sim 1.9$  \\
$L_{\rm Jeans}^{turb}$ (pc):        &  $\sim0.4$     &  $\sim 0.32$ \\
$M_{\rm Jeans}^{\rm turb}$ (\msun): &  $\sim 180$    &  $\sim 130$   \\
\hline
$L_{\rm frag}^{\rm ob}$ (pc):          &  $\sim 0.7$    &  $\sim 0.12$ \\
$M_{\rm frag}^{\rm ob}$ (\msun):   &  $\sim400$--500 &   $\sim 115$ \\
\hline
\end{tabular}
}

\end{table}

We carry out a virial analysis to assess the gravitational stability of dense cores.
Following the framework of \citet{Ber92}, the virial mass can be defined as
\begin{equation}
M_{\rm vir} = \frac{5 \sigma_{\rm tot}^{2} R_{\rm core}}{G},
\label{eq:viriall-mass}
\end{equation}
where $R_{\rm core}$ is the
core radius, and $G$ is the gravitational constant. 
Following Eq.\,2 of \citet{Liu19},
the total  velocity dispersion, $\sigma_{\rm tot}$, was calculated as $\sigma_{\rm tot}^2 = \sigma_{\rm th}^2 + \sigma_{\rm nt}^2$, 
to include both thermal and non-thermal support against gravity.
In this calculation, the same temperatures as assumed for calculating core masses were taken to be the kinetic temperature of the cores, while the line widths, $\Delta V_{\rm H^{13}CO^+}$, derived from the spectrum of \htcop~(1-0) averaged over each core (see Table\,\ref{tab:cores}) were
used for the $\sigma_{\rm nt}$ estimate.
The virial and  observed masses can be compared using the virial parameter,
\begin{equation}
\alpha_{\rm vir} = \frac{M_{\rm vir}}{M_{\rm core}} = \frac{{5}{\sigma_{\rm tot}^{2}}{R_{\rm core}}}{G M_{\rm core}}.
\label{eq:alpha}
\end{equation}
The significance of Eq.\,\ref{eq:alpha} is that supercritical cores with $\alpha_{\rm vir}\leq2$
will collapse towards star formation, while subcritical cores with $\alpha_{\rm vir}>2$ will expand or must be confined by additional forces (e.g., magnetic field and/or external pressure, \citealt{Kau13}).
As we see in Table\,\ref{tab:cores}, $\alpha_{\rm vir} < 2$ for all cores, which means that they are most likely gravitationally bound and will evolve to collapse, leading to star formation.

\subsection{Jeans length and Jeans mass \label{subsec:result:jeans}}

Figure\,\ref{fig:overview} clearly shows fragmentation on two different scales of the cloud and clumps.
On the cloud scale, the fragmentation proceeds with $\sim 0.7$\,pc-separated fragments of masses 200--500\,\msun\ (clumps MM1, MM2, and MM4 in Fig.\ref{fig:overview}a), where the fragment masses are  from Table\,1 of \citet{Liu20a} and the fragment separation is the averaged distance between neighbouring clumps. 
On the core scale, the fragmentation is evidenced by $\sim 0.12$\,pc-separated fragments of typical mass 115\,\msun, where only the fragments (cores) within MM1 are considered, since the population of cores in MM2 suffers from the incomplete coverage in the ATOMS.
In this case, the typical mass of the fragments is taken to be the average mass of the cores within MM1, while the typical separation between fragments is determined from the minimum spanning tree technique \egcite{Dib19}, which determines the shortest distances that can possibly connect each of the cores in the sampled field.

To investigate the observed two-scale fragmentation in the \filname\ cloud we evaluate the Jeans parameters, i.e., Jeans length and Jeans mass, of both cloud and clump scales \egcite{Wan14,Pal14}:
\begin{equation}
L_\mathrm{Jeans}= \sqrt{\frac{\pi c_\mathrm{eff}^2}{G\rho_\mathrm{eff}}},
\label{eq:jeans-len}
\end{equation}
and
\begin{equation}
M_\mathrm{Jeans}=  \frac{\pi^{5/2}}{6\,G^{3/2}}\, c_\mathrm{eff}^{3}\,\rho_\mathrm{eff}^{-1/2}.
\end{equation}
where $c_\mathrm{eff}$ and $\rho_\mathrm{eff}$ are the effective sound speed, and density, respectively.
If thermal support alone is considered, $c_\mathrm{eff}$ will be the isothermal sound speed (or the thermal velocity dispersion, $\sigma_\mathrm{th}$), and accordingly we obtain the thermal Jeans parameters, 
$L_\mathrm{Jeans}^{\rm th}$ and $M_\mathrm{Jeans}^{\rm th}$. 
If both thermal and non-thermal support are involved, $c_\mathrm{eff}$
will correspond to the total velocity dispersion $\sigma_\mathrm{tot}$ = $\sqrt{\sigma_{\rm th}^2 + \sigma_{\rm nt}^2}$, yielding the turbulent Jeans parameters, $L_\mathrm{Jeans}^{\rm turb}$ and $M_\mathrm{Jeans}^{\rm turb}$.

For the estimate of Jeans parameters on the cloud scale, we assume a gas kinetic temperature of $\sim 21$\,K, which corresponds to the average dust temperature over the entire cloud region investigated here (see Sect.\,\ref{subsec:result:cont}), and a line width of $\Delta\,V_{\rm H^{13}CO^{+}}=1.3$\,\vel, which is a typical value of the weak-feedback cloud area (see Fig.\,\ref{fig:kin:maps}c and Sect.\,\ref{subsec:result:molecular}), and thus more representative
of the initial turbulence of the cloud than the line widths found in the relatively strong-feedback areas. 
Following \citet{Liu19}, $\sigma_\mathrm{th}$ is estimated from 
the gas kinetic temperature and $\sigma_{\rm nt}$ is determined from $\Delta\,V_{\rm H^{13}CO^{+}}$. Finally, using the cloud-average density, $n_{\rm cl}\sim2.6\times10^5$\,\pcmcu\ (i.e., $\rho_{\rm eff}^{\rm cl}\sim4.4\times10^{-19}$\,g\,\pcmcu) both thermal and turbulent Jeans parameters are calculated on the cloud scale. 
Likewise, both thermal and turbulent Jeans parameters on the clump scale can be estimated assuming the same values of gas temperature and line width values as above.
This assumption is valid if both cloud- and clump-scale fragmentation occur sufficiently early that the cloud is free of strong star-formation feedback.
In this case, the mean density of the clump,
$n_{\rm clump}\sim4.9\times10^5$\,\pcmcu\ (\citealt{Liu20a}, i.e., $\rho_{\rm eff}^{\rm clump}\sim8.2\times10^{-19}$\,g\,\pcmcu) is adopted.

Table\,\ref{tab:fragment} summarizes the derived Jeans parameters at both cloud and clump scales.
At the cloud level fragmentation, the predicted thermal Jeans length is a factor of $\sim$7 smaller than the observed separation of the fragments (clumps) and the thermal Jeans mass is a factor of $\sim$150 - 200 smaller than the observed masses. In comparison, the turbulent Jeans parameters are closer to the observed values. Now, considering the fragmentation at the clump scale, the typical core separation observed is similar (within a factor of $\sim$2) to the predicted Jeans parameters with and without turbulence. In contrast, the predicted thermal Jeans mass is significantly (by a factor of $\sim$60) smaller than the observed core masses which are closer to the estimated turbulent Jeans mass. This strongly suggests that the observed hierarchical fragmentation seen at two different spatial scales of the \filname\ cloud is driven in part by turbulence.

As magnetic fields provide support against gravity, their role in the fragmentation process has been extensively investigated in several theoretical and simulation studies \citep[and references therein]{Com11,Pal13}. Recent observations by \citet{Pal21} of a sample of 18 fragmenting massive cores enable a comprehensive study of fragmentation and magnetic fields, where a tentative correlation between the fragmentation level and the magnetic field strength has been found with stronger magnetic fields corresponding to a less fragmentation level. From thermal dust polarization observations at 350~$\mu m$ within the filamentary \filname\ cloud, \citet{Tan19} propose a combination of gravity, turbulence, and magnetic field, with varying degree of contribution, to explain different levels of fragmentation in the associated clumps, MM1, MM2, and MM3. Based on SMA and CARMA observations \citep{Zha14,Hul14}, where no smaller scale structures are resolved in MM1, these authors conclude that the absence of observed fragmentation in this clump is due to the dominance of gravity over magnetic field and turbulence which enables a global collapse. In another study, \citet{Soa19} have mapped the magnetic fields in the \filname\ cloud at 870~$\mu m$ and estimated the field strength of the MM1 clump to be $\sim 500$\,$\mu$G, which would dominate given the inferred sub-alfv\'enic nature of the cloud. 

Our ATOMS continuum observations give a new insight into the above processes in MM1 by revealing a cluster of seven cores. In addition to the turbulence  driven fragmentation picture observed in \filname, the above analysis based on the previous studies suggests that magnetic fields play a decisive role in fragmentation both at the cloud and the clump scale, especially for the fragmentation discerned in the clump MM1. Comparing the distribution of the resolved cores with the orientation of the magnetic field given in \citealt{Soa19, Tan19} indicates that fragmentation has ensued  mostly along a preferred direction perpendicular to the magnetic field. This agrees well with strong magnetic field cases discussed in \citet{Pal21}.

Additionally,  
recent statistical studies towards several tens of high-mass star-forming protoclusters or IRDC clumps with interferometic observations down to $\sim 1000$\,AU scales \egcite{Pal13, Beu18, San19}, have revealed large populations of low-mass cores which are consistent with thermal fragmentation. 
As can be deciphered from these studies and the analysis carried out in this present work, massive fragments (at the scale of clump, cores, or condensations) are additionally supported by turbulence and/or magnetic field where the core/condensations eventually form high-mass stars. In case of small scale condensations, which would either competitively accrete to form high-mass stars or proceed to form the low-mass population of the protocluster, thermal pressure dominates and these are consistent with thermal Jeans fragmentation.  
Further high-resolution observations of the ATOMS cores would enable us to probe the fragmentation process involved from core to condensation (or star-forming `seed') scales in the \filname\ cloud. In this regard, surveys like ALMAGAL hold the potential for in-depth studies of a large sample of similar star-forming regions to gain a better insight into the complex fragmentation processes involved. In summary, while the fragmentation details on the core scales require future in-depth studies, the current observations from the ATOMS suggest that turbulence and magnetic field could act together to drive the fragmentation of both cloud and clump scales.

\section{Discussion}
\subsection{Hierarchical fragmentation\label{subsec:disc:frag}}
It is widely accepted that hierarchical fragmentation takes place in the star formation process on multiple scales, from molecular cloud to clump, and down to core scales.
As shown in Fig.\,\ref{fig:overview}, the \filname\ cloud does present a hierarchical fragmentation scenario with the 
intermediate-scale clumps fragmented from the large-scale filamentary cloud, followed by the small-scale cores fragmented from the
clumps.
Moreover, subsequent fragmentation  of these detected cores is likely to occur as has been observed towards other star-forming clumps in previous high-resolution observations \egcite{Pal13,Beu18}, where the fragmentation on core scales 
into $<1000$\,AU-size fragments has been revealed.

The observed fragmentation scenario in \filname\ could favour the hierarchical fragmentation-based models such as the ``global hierarchical collapse'' and ``inertial-inflow'' models \citep{Vaz19,Pad20}.
The former model assumes that fragmentation can occur hierarchically on all scales through the thermal Jeans fragmentation of the relative-scale density structures being in transonic and/or subsonic state.  In contrast, the latter model of ``inertial-inflow''  \citep{Pad20} requires supersonic turbulence to trigger the hierarchical fragmentation, which is consistent with the fragmentation analysis carried out in Sect.\,\ref{subsec:result:jeans}. 
 Certainly, the understanding of the exact driving agents, thermal versus turbulent fragmentation, in the hierarchical fragmentation-based models needs to be constrained in future studies. Moreover, from
 a theoretical point of view the above models  could be complementary to the class of other high-mass star formation models like the ``core-accretion'' and ``competitive accretion'' models \citep{McK03,Bon04}. For example, the ``global hierarchical collapse'' model claims that the ``Bondi-Hoyle'' accretion still holds on core scales as predicted by the ``competitive accretion'' model.
To reconcile these two classes of theoretical models requires  high-resolution observations of both continuum and lines on multiple scales, i.e., from a few parsecs to a few hundred AUs. 
This strongly advocates for future higher-resolution, and sufficiently deep observations to scrutinize the core-scale fragmentation process of the \filname\ cloud.

\subsection{Dynamical mass inflow and accretion \label{subsec:disc:dynamic}}
The multi-scale, hierarchical fragmentation process could coexist with rich, and characteristic dynamics, such as scale-dependent
gas flows and mass accretion \egcite{Per13,Yua18,Avi21,Ren21}. 
Such dynamical processes are believed to determine the final mass of the newly-formed stars \egcite{Beu18,Mot18,Vaz19,Pad20}. 
Therefore, it is worthwhile to investigate the dynamics of \filname\ to gain further insight into the fragmentation scenario of the cloud.

On the clump scale, we find from Fig.\,\ref{fig:kin:maps}b the evident large-scale  velocity-coherent gradients along several directions, for example A and B, towards the cluster of cores within MM2.
These velocity gradients are reliable since they are measured from \htcop~(1--0), whose emission has a single-peak spectrum almost everywhere especially for the regions investigated here. They are estimated to be around a few \vel\,pc$^{-1}$ (see Fig.\,\ref{fig:vel:grad}). 
Both outflowing and inflowing gas can contribute to the apparent velocity-coherent gradients. 
Moreover, we do not find the outflow signatures (i.e., enhanced, high-speed blue or red-shifted emission) from the visual inspection of the PV diagram of \htcop~(1--0) at several particularly selected directions (in Fig.\,\ref{fig:kin:maps}b). These results indicate that the relatively optically thin \htcop~(1--0) emission is not significantly affected by outflows.

As mentioned in Sect.\ref{subsec:result:molecular}, the bright branch of the \htcop\ emission (see Fig.\,\ref{fig:kin:maps}a), being morphologically linked to the cluster of cores in MM2, could be an imprint of ambient gas inflowing onto the cluster due to its strong gravitational attraction. 
In support of this picture, the imprint of the inflowing ambient gas is suggested by a spoke-like gas streamers converging toward the centre of the cluster in MM2 from ${\rm N_2H^+}$~(1--0) emission of the ALMA-IRDC survey \citep{Bar21} that reaches a matching angular resolution to the ATOMS survey.  Analysing the ${\rm N_2H^+}$~(1--0) gas kinematics towards \filname\ , \cite{Tan19} have discussed about the presence of a large-scale east-west velocity gradient. At smaller and localized scales towards the MM1/MM2 clumps, close alignment between the local magnetic field orientation is also evident indicating that gravity has aligned the gas flow along the field lines onto the protoclusters MM1/MM2. These results agree well with the observed velocity-coherent gradients illustrated in Fig.\,\ref{fig:kin:maps}b.

 For a quantitative analysis, we proceed along the lines discussed in \citet{Mos21} to estimate the mass inflowing rate of the gas flows onto the cluster of cores within both MM1 and MM2. Based on the peak intensity map (see Fig.\,\ref{fig:kin:maps}a), we approximately delineate the gas inflow regions with two polygons (see Fig.\,\ref{fig:kin:maps}b),
each morphologically encompassing the gas emission possibly linked to each cluster of cores.
Here, the mass inflowing rate, $\dot M_{\rm inf}$, is calculated from the ratio 
of the momentum, $P_{\rm inf}$, to the length, $L_{\rm inf}$, of the flow,  similar to the approach followed in
calculating the outflow properties in Sect.\,\ref{subsec:result:outflowsMM1}.
$L_{\rm inf}$ corresponds to the largest extent of the gas inflowing areas (see Fig.\,\ref{fig:kin:maps}b), i.e.,  $\sim 0.26$\,pc and $\sim 0.33$\,pc for the cluster of cores in MM1, and MM2, respectively.
 We assume a
clump-averaged abundance of $9\times10^{-12}$  for \htcop~(1-0) as estimated in \citet{Liu20a}.
The velocities of inflowing gas are in the range [53, 61]\,\vel\ bracketing the systemic velocity of $57.6$\,\vel\ for the individual MM1 system, and $57$\,\vel\ for MM2.
Following Sect.\,\ref{subsec:result:outflowsMM1}, and confining the analysis to the gas inflowing area for each cluster of cores, we  find that the gas could be inflowing onto the cluster of cores in MM1 at $\dot M_{\rm inf}\simeq 2.7\times10^{-4}$\,\msun~yr$^{-1}$, and in MM2 at $\dot M_{\rm inf}\simeq3.6\times10^{-4}$\,\msun~yr$^{-1}$.
These results are in good agreement with those of \citet{Mos21}, who observed the mass inflow rate onto the core cluster to be around $10^{-4}$\,\msun~yr$^{-1}$ in a high-mass star-forming clump in IRAS 21078+5211.
On the core scale, several associated, highly-collimated outflows are found in both clusters of cores. 
The outflows are generally thought to be an indirect evidence for disk accretion.

$\dot M_{\rm inf}$ corresponds to the mass inflow of clump scale gas onto the cores, while $\dot M_{\rm acc}$ corresponds to the mass accretion of the core scale onto the stellar disk. 
It is therefore natural to compare 
the mass inflow/accretion rates of different scales.
We find that $\dot M_{\rm inf} >> \dot M_{\rm acc}$, suggesting that the mass inflow/accretion rate from the clump to core scales is 
higher than that from the core to  disk scales. 
While the observed scale-dependent mass inflow/accretion rates seem to agree well with the ``inertial-inflow" model \citep{Pad20}, one cannot rule out the ``global hierarchical collapse" scenario. In the ``inertial-inflow" model, the mass inflow/accretion rate is predicted to be a growing function of distance from the centre of star-forming clusters, so that the large-scale mass inflow rate can control the small-scale mass accretion rate onto the star(s), 
leading to a cascade of the scale-dependent mass feeding
from large to small scales. In comparison, the scenario of the large scale density structures regulating the small-scale structures in dynamical accretions is also predicted in the ``global-hierarchical collapse'' model which suggests a similar cascade of scale-dependent mass feeding due to the top-down mass accretion process from large to small scales \citep{Vaz19}. Therefore, further dedicated dynamics and kinematics analysis in the future remains necessary to distinguish between the two models from an observational point of view.

This cascade of scale-dependent mass feeding scenario 
has already been demonstrated from observations \egcite{Mot18,Yua18,Avi21}, 
in which the central protostar, the core, and the clump are found to simultaneously grow in mass via core-fed/disk accretion, clump-fed accretion, and filamentary/cloud-fed accretion respectively, with a trend of increasing mass inflow/accretion rate. This suggests that high-mass star formation could be a dynamical mass inflow/accretion process linked to the multi-scale fragments from the clouds, through clumps and cores, down to seeds or condensations of star formation.
We therefore suggest that the high-mass star-forming \filname\ cloud could be undergoing a multi-scale, and dynamical inflow/accretion process, which could be linked to multi-scale fragmentation.

\section{Summary and conclusions} \label{sec:summary}
We have presented new observations of 3\,mm continuum and  molecular transitions (e.g.,  HCO$^+$/H$^{13}$CO$^+$ $J=$1--0, and CS $J=$2--1 from the ATOMS survey) for tracing both dense gas and outflows in the two massive protostellar clumps, MM1 and MM2, in the \filname\ filamentary IRDC cloud. 
We have analyzed the fragmentation and dynamics of the cloud down to the cores of $\sim 0.03$\,pc in radius, and our main results are the following:
\begin{itemize}
	\item Nine dust cores have been extracted from 3\,mm dust continuum emission:  seven within the MM1 clump and two within MM2. 
	The seven cores represent the most complete population of cores unveiled so far for MM1, while the small population of only two detected cores in MM2 is mainly attributed to the incomplete coverage by the ATOMS survey.
	\item The nine cores have median radius 0.03\,pc (range 0.02--0.04\,pc), median mass 115\,\msun\ (range $\sim$~40--280\,\msun), median number density $1.1\times10^{7}$\,\pcmcu\ (range 0.2$\times10^{7}$--2.8$\times10^{7}$\,\pcmcu), and median mass surface density 11\,g\,\pcmsq\ (range $\sim$2--33\,g\,\pcmsq).
	All of the cores have viral parameter $\alpha_{\rm vir}<2$, suggesting that they are most likely gravitationally bound and will proceed towards the formation of new stars.
	\item The identification of new outflows suggests the presence of at least four individual, highly-collimated outflows within MM1 as opposed to the two wide-angle outflows reported in \citet{She07}. 
	The outflow properties in MM1 are confirmed to be attributed to a B0-type star with a total outflowing mass of 
	$\sim 45$\,\msun, and total energy of $\sim 1\times10^{47}$\,ergs. Additionally, a total mass accretion rate onto the protostars in MM1, $\dot M_{\rm acc}$, is
	estimated to be of the order of $10^{-5}$\,\msun\,yr$^{-1}$ in the framework of momentum conservation between mass infall and momentum-driven outflows in a protostellar-jet/outflow system.
	\item A large-scale, butterfly-shaped velocity gradient pattern is observed in \htcop~(1--0) emission surrounding the MM1-a core, which is consistent with the picture revealed in \chthoh\ at $345$\,GHz by \citealt{Rat11}.
	The large-scale nature of the pattern (size $\sim$ 0.18\,pc) suggests that the observed velocity gradient could arise from a large, rotating structure rather than from a small, rotating disk.
	\item Two-scale hierarchical fragmentation is evident on both cloud and clump scales. It could be driven by a combination of the initial turbulence and magnetic field on both scales.
	With the potential of the cores to fragment further into star formation seeds of sizes $<\sim$ 1000 AU, we assume that a multi-scale, hierarchical fragmentation process
	is at work in the \filname\ cloud.
	\item Intermediate-scale velocity gradients towards each cluster of cores are found in both MM1 and MM2 clumps with a typical amplitude of 3--8\,\vel~pc$^{-1}$. 
	These are interpreted as the gas inflowing onto the cluster in the context of the multi-scale, hierarchical fragmentation.
	The corresponding mass inflow rate, $\dot M_{\rm inf}$, is estimated to be of order of $10^{-4}$\,\msun\,yr$^{-1}$.
	\item $\dot M_{\rm acc}$ responsible for the small-scale mass accretion from cores onto protostars and disks is found to be lower than $\dot M_{\rm inf}$ for the larger-scale mass inflow from clumps onto the cluster of cores. 
	This difference suggests a scale-dependent inflow/accretion cascade scenario from large to small scales, which could be linked to the multi-scale fragmentation process.

\end{itemize}

    Multi-scale fragmentation from clouds, through clumps and cores, down to seeds of star formation and the cascade of scale-dependent mass inflow/accretion observed in \filname\ allow us to conclude that the cloud is undergoing a dynamical mass inflow/accretion process. 
    Confirmation of this process would support hierarchical fragmentation-based models,  e.g., ``global hierarchical collapse'' 
    and ``inertial-inflow'' models.

\medskip
\noindent{\textbf{Acknowledgements}}\\
We thank the anonymous referee for comments and suggestions that greatly improved the quality of this paper.
Tie Liu acknowledges the supports by National Natural Science Foundation of China (NSFC) through grants No.12073061 and No.12122307, the international partnership program of Chinese academy of sciences through grant No.114231KYSB20200009, and Shanghai Pujiang Program 20PJ1415500.
This research was carried out in part at the Jet Propulsion Laboratory, which is operated by the California Institute of Technology under a contract with the National Aeronautics and Space Administration
(80NM0018D0004).
S.-L. Qin is supported by NSFC under No.12033005.
L.B. and A.S. acknowledge support from CONICYT project Basal AFB-170002. AS gratefully acknowledges funding support through Fondecyt Regular (project code
1180350). 
C.W.L. is supported by Basic Science Research Program through the National Research Foundation of Korea (NRF)
funded by the Ministry of Education, Science and Technology
(NRF-2019R1A2C1010851).
This paper makes use of the following ALMA data: ADS/JAO.ALMA\#2019.1.00685.S. ALMA is a partnership of ESO (representing its member states), NSF (USA), 
and NINS (Japan), together with NRC (Canada), MOST and ASIAA (Taiwan), and KASI (Republic of Korea), in cooperation with the Republic of Chile. The Joint 
ALMA Observatory is operated by ESO, AUI/NRAO, and NAOJ.
This research made use of astrodendro, a Python package to compute dendrograms of Astronomical data ({\url{http://www.dendrograms.org/}}).
This research made use of Astropy,
a community-developed core Python package for Astronomy (Astropy
Collaboration, 2018).

\noindent{\textbf{Data availability}}\\
The data underlying this article are available in the article and in its online supplementary material.

\vspace{-5mm}

%\clearpage

%\begin{thebibliography}{}
\input ourwork.bbl
%\end{thebibliography}
% 
%\clearpage
\appendix
\section{Outflows seen in other molecules}
%\label{sec:app-kin-dist} 

\begin{figure*}
\centering
\includegraphics[width=6.8 in]{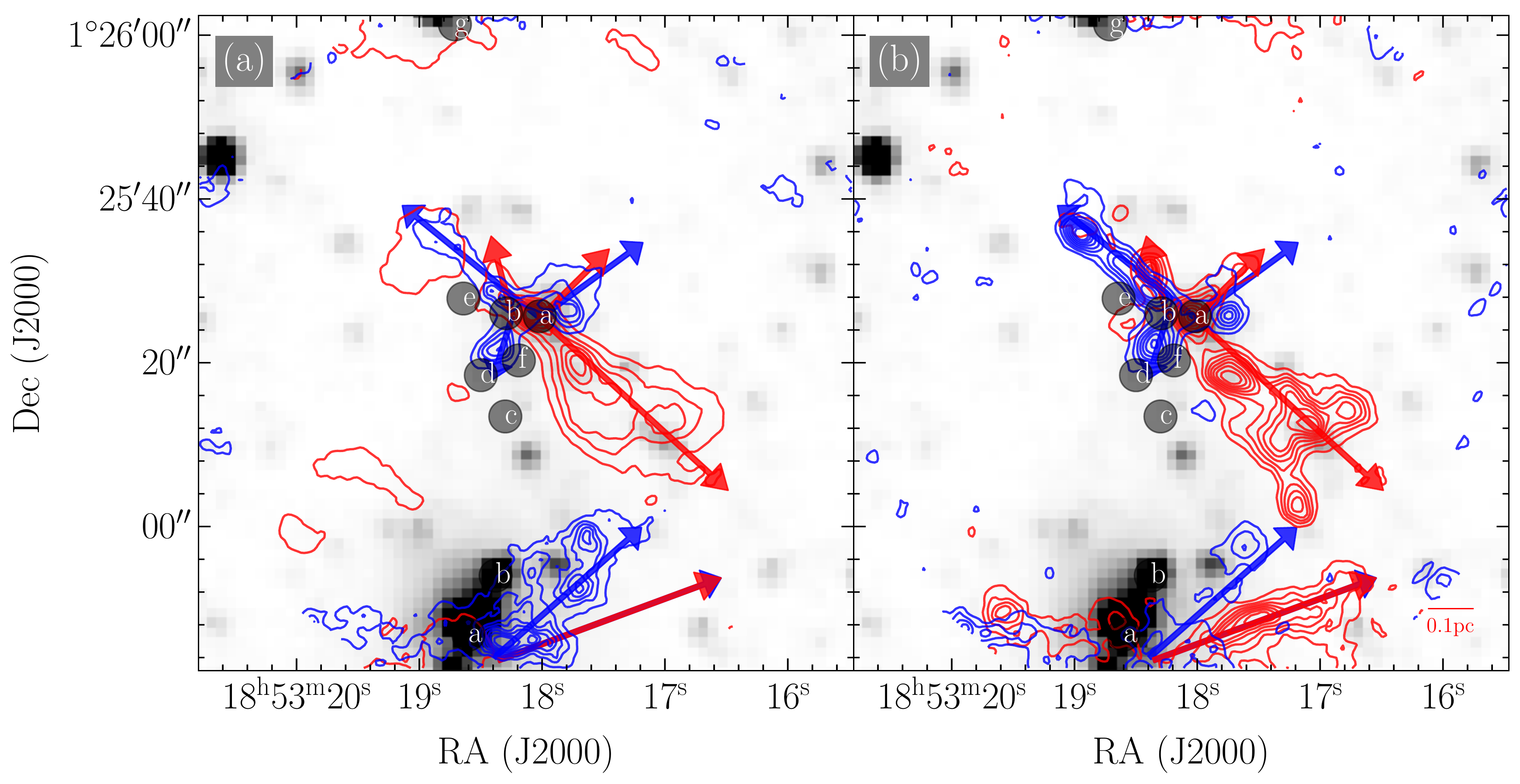}
\includegraphics[width=6.8 in]{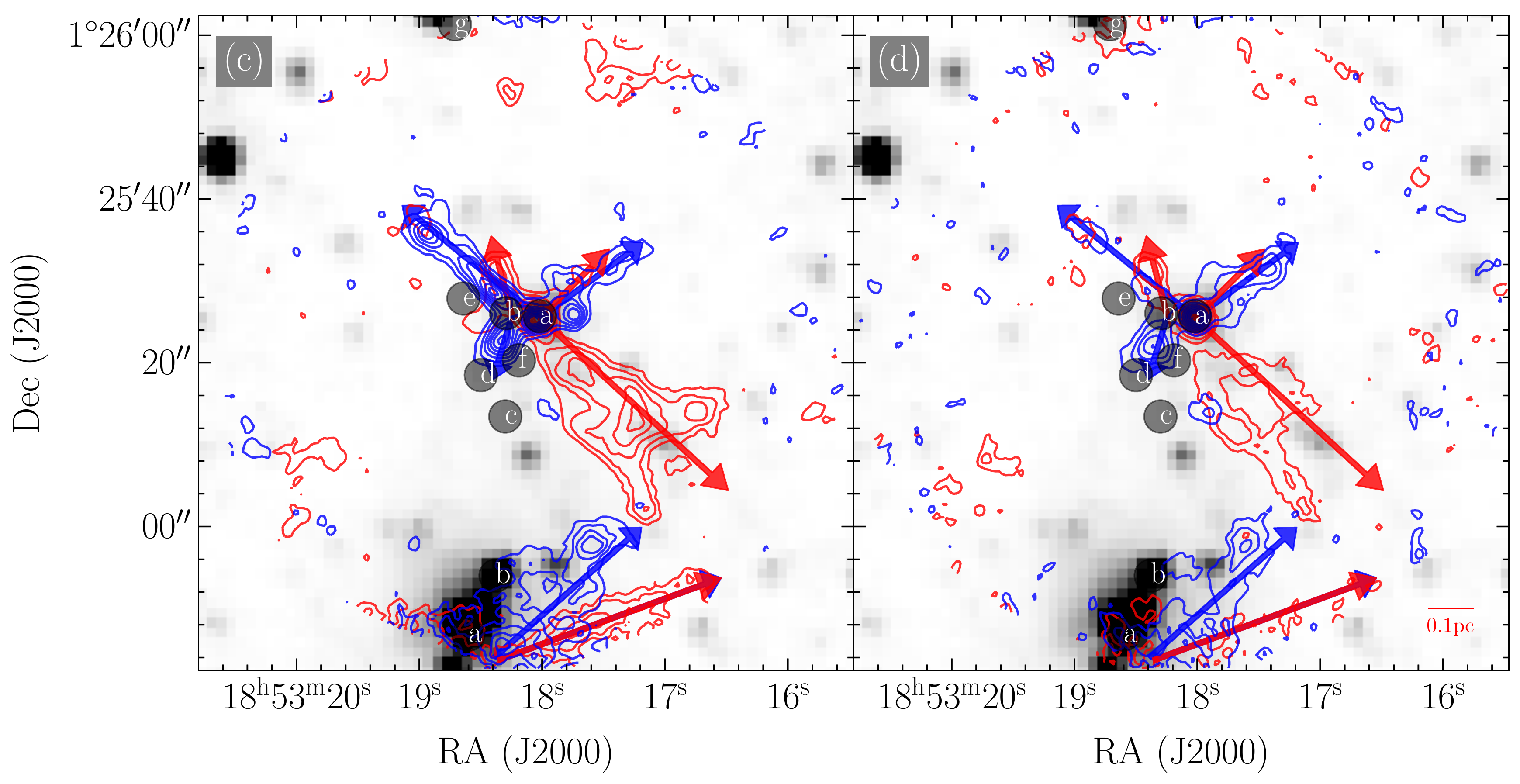}
\caption{Outflow emission traced by different tracers overlaid on the {\it Spitzer} 4.5\,$\mu$m) image. 
Blue- and red-shifted outflowing gas emission in order is integrated  over [44.0, 55.0]\,\vel\ and [60.0, 76.0]\,\vel\ for \hcop~(1-0) in panel\,a,
[42.0, 51.3]\,\vel\ and [63.5, 74.0]\,\vel\ for SiO in panel\,b,  [42.0, 52.7]\,\vel\ and [62.2, 78.0]\,\vel\ for SO in panel\,c, 
and [46.0, 54.2]\,\vel\ and [60.6, 66.0]\,\vel\ for \chthoh\ in panel\,d. 
Labels\,a--g identify the dense cores in both MM1 and MM2 protostellar clumps.
The red and blue arrows correspond to the axes of the CS~(2-1) outflow lobes.
}
\label{fig:outflow:gallary}
\end{figure*}

%\clearpage
%\onecolumn

\vspace{5mm}
\noindent
Author affiliations:\\

\noindent 
$^{1}$Department of Astronomy, Yunnan University, Kunming, 650091, PR China \\
$^{2}$Indian Institute of Space Science and Technology, Thiruvananthapuram 695 547, Kerala, India\\
$^{3}$Shanghai Astronomical Observatory, Chinese Academy of Sciences, 80 Nandan Road, Shanghai 200030, Peoples Republic of China \\
$^{4}$Key Laboratory for Research in Galaxies and Cosmology, Shanghai Astronomical Observatory, Chinese Academy of Sciences, 80 Nandan Road, Shanghai 200030, Peoples Republic of China \\
$^{5}$Jet Propulsion Laboratory, California Institute of Technology, 4800 Oak Grove Drive, Pasadena, CA 91109, USA\\
$^{6}$Key Laboratory of Radio Astronomy, Chinese Academy of Sciences, Nanjing 210008, People's Republic of China\\
$^{7}$Center for Astrophysics $|$ Harvard \& Smithsonian, 60 Garden Street, Cambridge, MA 02138, USA\\
$^{8}$Kavli Institute for Astronomy and Astrophysics, Peking University, 5 Yiheyuan Road, Haidian District, Beijing 100871, People's Republic of China\\
$^{9}$Department of Astronomy, Peking University, 100871, Beijing, People's Republic of China\\
$^{10}$Korea Astronomy and Space Science Institute, 776 Daedeokdaero, Yuseong-gu, Daejeon 34055, Republic of Korea\\
$^{11}$SOFIA Science Centre, USRA, NASA Ames Research Centre, MS-12, N232, Moffett Field, CA 94035, USA \\
$^{12}$Physical Research Laboratory, Navrangpura, Ahmedabad—380 009, India \\
$^{13}$University of Science and Technology, Korea (UST), 217 Gajeong-ro, Yuseong-gu, Daejeon 34113, Republic of Korea\\
$^{14}$Astronomy Department, University of California, Berkeley, CA 94720, USA\\
$^{15}$School of Physics and Astronomy, Sun Yat-sen University, 2 Daxue Road, Zhuhai, Guangdong, 519082, People's Republic of China\\
$^{16}$National Astronomical Observatories, Chinese Academy of Sciences, Beijing 100101, China  \\
$^{17}$Department of Physics, P.O. box 64, FI- 00014, University of Helsinki, Finland \\
$^{18}$Departamento de Astronom\'{\i}a, Universidad de Chile, Las Condes, Santiago, Chile\\
$^{19}$National Astronomical Observatory of Japan, National Institutes of Natural Sciences, 2-21-1 Osawa, Mitaka, Tokyo 181-8588, Japan\\
$^{20}$Center for Astrophysics, GuangZhou University, Guangzhou, China\\
$^{21}$University of Chinese Academy of Sciences, Beijing 100049, China\\
$^{22}$NAOC-UKZN Computational Astrophysics Centre, University of KwaZulu-Natal, Durban 4000, South Africa\\
$^{23}$Departamento de Astronom\'ia, Universidad de Concepci\'on, Av. Esteban Iturra s/n, Distrito Universitario, 160-C, Chile \\
$^{24}$Max-Planck-Institute for Astronomy, K\"{o}nigstuhl 17, 69117 Heidelberg, Germany \\
$^{25}$E\"{o}tv\"{o}s Lor\'{a}nd University, Department of Astronomy, P\'{a}zm\'{a}ny P\'{e}ter s\'{e}t\'{a}ny 1/A, H-1117, Budapest, Hungary\\
$^{26}$Department of Physics, Taiyuan Normal University, Jinzhong 030619, China \\
$^{27}$Satyendra Nath Bose National Centre for Basic Sciences, Block-JD, Sector-III, Salt Lake, Kolkata-700 106 \\
$^{28}$Indian Institute of Science Education and Research Tirupati, Rami Reddy Nagar, Karakambadi Road, Mangalam (P.O.), Tirupati 517 507, India\\

% \bsp
\label{lastpage}
\end{document}